\documentclass[aps,preprint,nofootinbib,eqsecnum]{revtex4}%
\usepackage{amsmath}
\usepackage{graphicx}
\usepackage{amsfonts}
\usepackage{amssymb}%
\providecommand{\U}[1]{\protect\rule{.1in}{.1in}}
\providecommand{\U}[1]{\protect\rule{.1in}{.1in}}
\providecommand{\U}[1]{\protect\rule{.1in}{.1in}}

\begin{document}

\begin{center}

{\leftline {USC-07/HEP-B4 \hfill arXiv:0705.2834 [hep-th]}}

{\vskip0.8cm}

{\Large Dual Field Theories In $(d-1)+1$ Emergent Spacetimes }

{\Large From A Unifying Field Theory In $d+2$ Spacetime}\footnote{This work
was partially supported by the US Department of Energy, grant number
DE-FG03-84ER40168.}

{\vskip0.3cm}

\textbf{Itzhak Bars, Shih-Hung Chen and Guillaume Qu\'{e}lin}

{\vskip0.2cm}

\textsl{Department of Physics and Astronomy, }

\textsl{University of Southern California, Los Angeles, CA 90089-0484, USA.}

{\vskip0.8cm}

\textbf{Abstract}
\end{center}

According to Two-Time Physics, there is more to space-time than can
be garnered with the ordinary formulation of physics. Two-Time
Physics has shown that the Standard Model of Particles and Forces is
successfully reproduced by a two-time field theory in 4 space and 2
time dimensions projected as a holographic image on an emergent
spacetime in 3+1 dimensions. Among the successes of this approach is
the resolution of the strong CP problem of QCD as an outcome of the
restrictions imposed by the higher symmetry structures in 4+2
dimensions. In this paper we launch a program to construct the duals
of the Standard Model as other holographic images of the same 4+2
dimensional theory on a variety of emergent spacetimes in 3+1
dimensions. These dual field theories are obtained as a family of
gauge choices in the master 4+2 field theory. In the present paper
we deal with some of the simpler gauge choices which lead to
interacting Klein-Gordon field theories for the conformal scalar
with a predicted SO(d,2) symmetry in a variety of interesting curved
spacetimes in (d-1)+1 dimensions. More challenging and more
interesting gauge choices (including some that relate to mass) which
are left to future work are also outlined. Through this approach we
discover a new realm of previously unexplored dualities and hidden
symmetries that exist both in the macroscopic and microscopic
worlds, at the classical and quantum levels. Such phenomena
predicted by 2T-physics can in principle be confirmed both by theory
and experiment. 1T-physics can be used to analyze the predictions
but in most instances gives no clue that the predicted phenomena
exist in the first place. This point of view suggests a new paradigm
for the construction of a fundamental theory that is likely to
impact on the quest for unification.

\newpage

\tableofcontents
\newpage

\section{2T-physics versus 1T-physics}

Evidence has been mounting that the ordinary formulation of physics, in a
space-time with three space and one time dimensions (1T-physics), is
insufficient to describe certain aspects of our world, just like shadows on
walls alone are insufficient to capture the true essence of an object in a
three dimensional room. Two-Time Physics (2T-physics) \cite{2T basics}%
-\cite{spinBO} has revealed that our physical world in three space and one
time dimensions is like a holographic shadow of a highly symmetric universe in
four space and two time dimensions. Amazingly, the fundamental theory in
Physics that agrees exquisitely with experiment as we know it today, the
Standard Model of Particles and Forces in 3+1 dimensions, is reproduced, and
to boot its \textquotedblleft strong CP problem\textquotedblright\ is solved,
as a holographic \textquotedblleft shadow\textquotedblright\ of a field theory
in 4+2 dimensions \cite{2T SM}.

A surprising outcome of the 4+2 point of view (more generally $d+2$) is that
the master theory has many 3+1 dimensional holographic \textquotedblleft
shadows\textquotedblright\ (more generally $\left(  d-1\right)  +1$), which
are distinguishable from the point of view of 1T-physics. Thus, the 2T-physics
formulation predicts that the usual Standard Model has close relatives, i.e.
other \textquotedblleft shadows\textquotedblright,\ that look like different
field theories, but yet represent the same theory in 4+2 dimensions, so that
they are dual to the Standard Model. A variety of \textquotedblleft
shadows\textquotedblright\ are predicted to appear with different dynamics in
1T-physics (i.e. different Hamiltonians, or different field equations), to
have hidden symmetries that signal an extra space and an extra time, and to
have an infinite set of calculable and measurable relationships among them,
akin to dualities.

Since these \textquotedblleft shadows\textquotedblright\ belong to the same
theory, a new type of unification of different field theories emerges through
higher dimensions that includes one extra space and one extra time dimensions.
This type of unification is very different from the Kaluza-Klein idea because
there are no Kaluza-Klein type remnant degrees of freedom from the extra 1+1
spacetime, but instead there is a family of dual field theories.

The extra 1+1 dimensions required by 2T-physics are different than
Kaluza-Klein type of extra dimensions, however both types of extra dimensions
could coexist in the same theory since their roles are not incompatible with
each other.

The reader that is familiar with M theory may recognize some parallels to
dualities in M theory. In this paper we will make the case that there are
conceptually similar dualities at the level of field theory that include the
Standard Model.

In 2T-physics the concepts outlined above are realized in explicit dynamical
theories which apply to both macroscopic and microscopic physics, at the
classical and quantum levels. The dualities are related to a gauge symmetry
which treats the extra 1+1 space-time as gauge degrees of freedom embedded in
$d+2$ dimensions.

Such properties have previously been exhibited with explicit examples in the
worldline formalism for particles \cite{2T basics}, including spin
\cite{spin2t}\cite{spinBO}, supersymmetry \cite{super2t}\cite{2ttwistor}%
\cite{twistorLect}\cite{2tsuperstring}, and interactions with background
fields \cite{2tgravembackgrounds}\cite{2tbacgrounds}. What ties the dual
images together is the higher dimensional \textquotedblleft master
theory\textquotedblright\ in $d+2$ dimensions which is subject to certain
gauge symmetries. As the master theory is changed (and there is an infinite
set of possibilities in the worldline formalism), new shadows are generated
with new duality relationships and new hidden symmetries.

The underlying reason for such striking properties cannot be found in
1T-physics in $(d-1)+1$ dimensions, but is explained in 2T-physics \cite{2T
basics} as being due to a fundamental Sp$\left(  2,R\right)  $ gauge symmetry
which acts in \textit{phase space} $\left(  X^{M},P_{M}\right)  $ and makes
position and momentum indistinguishable at any instant. This gauge symmetry
can be implemented only if the target spacetime includes one extra space and
one extra time dimensions, thus showing that the unification relies on a
spacetime in $d+2$ dimensions demanded by the Sp$\left(  2,R\right)  $ gauge symmetry.

In general the $d+2$ dimensional spacetime is not necessarily flat, so
background fields of all spins are permitted provided they satisfy some
restrictions in $d+2$ dimensions \cite{2tgravembackgrounds}\cite{2tbacgrounds}.

The evidence of the $d+2$ dimensional world in the form of hidden symmetries
and dualities can be found both at the macroscopic and microscopic scales, and
such predictions of Two-Time Physics can be tested through both theory and
experiment. The existence of such relations and hidden symmetries, never
suspected in 1T-physics, shows that the higher space in $d+2$ dimensions is
not just formalism that could be avoided. 1T-physics can be used to verify and
interpret the predictions of 2T-physics, but it is not equipped to come up
with the predictions in the first place, unless one stumbles into some of them
occasionally, such as SO$\left(  4,2\right)  $ conformal symmetry of massless
systems. The lessons from 2T-physics so far makes it evident that the ordinary
1T-physics formulation of Nature is insufficient to provide the explanation or
even the existence of the many unifying facts revealed through 2T-physics.

To settle the physical and philosophical interpretation of 2T-physics will
probably require much discussion in the future. For the time being we are
content to use the 2T formalism at least as an approach that provides new
mathematical tools and new physical insights for understanding our universe.
In this paper we launch a program to exploit such properties of 2T-physics in
the context of field theory, to develop new analytical methods that would be
useful in their own right directly in 1T-physics, and that will help us
understand the deeper implications of 2T-physics.

\section{New Emergent Principles in Field Theory}

As mentioned above, two-time physics \cite{2T basics} is based on gauging the
symplectic transformations of phase space $\left(  X^{M},P_{M}\right)  $,
Sp$\left(  2,\mathbb{R}\right)  $. One of the fundamental results of this new
gauge principle is that, in order to be nontrivial, it requires the theory to
be formulated in a spacetime having at least two times. \ While the choice of
exactly two timelike dimensions results in a coherent theory, investigations
of alternatives with more than two times have been done \cite{3times}. So far,
these appear to rule out such possibilities and seem to confirm the special
status of 2T-physics. \

The theory was first formulated in the worldline formalism in which the
Sp$\left(  2,\mathbb{R}\right)  $ gauge symmetry allows the elimination of one
spacelike and two timelike degrees of freedom. When quantized, the theory is
seen to be completely free of ghosts, thus confirming the viability of the
theory. In this paper our investigation will be at the level of field theory,
rather than worldline theory. The connection between the two is that the field
$\Phi\left(  X\right)  $ corresponds to the first quantized wavefunction
$\Phi\left(  X\right)  =\langle X|\Phi\rangle$, and in the field theory
context we also include field interactions.

Although an infinite set of Sp$\left(  2,R\right)  $ gauge invariant theories
exist for describing particles on the worldline moving in arbitrary
backgrounds \cite{2tgravembackgrounds}\cite{2tbacgrounds}, so far most of the
investigations of 2T-physics have concentrated on the free particle in flat
$d+2$ dimensions with an SO$\left(  d,2\right)  $ global symmetry. This
simplest flat background in $d+2$ dimensions, which leads to a rich set of
backgrounds and dualities in $\left(  d-1\right)  +1$ dimensions, as discussed
in section \ref{gauge choices}, will also be adopted in the context of
2T-field theory in this paper.

An important general feature of 2T physics on the worldline is that the
potentially infinite variety of gauge choices leads to an equally infinite
family of lower dimensional systems. All the possible gauge choices have not
been classified. A list of the known gauge choices of the simplest theory on
the worldline is provided in section \ref{gauge choices}. It should be noted
that, while the master worldline theory is the free particle in flat spacetime
in $d+2$ dimensions subject to the Sp$\left(  2,R\right)  $ constraints, the
emergent $\left(  d-1\right)  +1$ dimensional worldline systems include both
free and interacting particles, in flat or curved spaces, with or without
mass. The parameters that describe mass, interaction, curvature, etc. emerge
from the extra dimensions as moduli that parameterize the gauge fixed phase space.

The issue of gauge choice is related to the question of which $\left(
d-1\right)  +1$ dimensions is embedded in $d+2$ dimensions. The different
embeddings of \textit{phase space} (not just space) in $\left(  d-1\right)
+1$ dimensions into phase space in $d+2$ dimensions potentially creates a huge
variety of dual field theories. It is expected that the investigation of these
duals in the context of field theory could lead to further insight into
nonperturbative aspects of the theory (such as QCD) and might also provide new
calculational tools.

While these systems look very different in the particle worldline theory
(different Hamiltonians), the higher dimensional theory actually proves that
these are all dual to one another and establishes that dualities must connect
them - a duality which would have been hard to show otherwise. \ In
particular, all these systems have a hidden global SO$\left(  d,2\right)  $
symmetry which is a manifest global symmetry of the parent theory in flat
space. \ The different realizations of this symmetry in the lower dimensions
are often highly nonlinear and, again, would have been hard to find directly.
\ The gauge choices leading to different 1T-physics systems have their
counterparts in field theory, and the construction of the corresponding dual
field theories, and their hidden SO$\left(  d,2\right)  $ global symmetry, is
the focus of our investigation.\

The worldline gauge choices provided in section \ref{gauge choices} will be
adopted for similar gauge choices in 2T-physics field theory that will be
discussed in this paper. These will lead to the \textit{emergent field
theories in various spacetimes}. In addition to the interaction features and
parameters of the emergent spacetimes mentioned above (mass, interaction,
curvature, etc.), the 2T-physics field theory adds local field interactions in
$d+2$ dimensions. Thus a given emergent field theory system in $\left(
d-1\right)  +1$ dimensions, which is a member of the duality family, receives
contributions to its interaction structure both from the embedding in $d+2$
dimensions and from the field interactions directly in $d+2$ dimensions.

First quantization of the emerging worldline systems in specific gauges can be
tricky in terms of the respective phase spaces, since in first quantization
ordering issues of nonlinear expressions must be taken care of in order to
preserve the SO$\left(  d,2\right)  $ symmetry at the quantum level. \ A
recent success that overcomes this issue automatically was the formulation of
the field theory approach directly at the two-time level, with its own new
gauge symmetry that is related to the underlying Sp$\left(  2,R\right)  $
\cite{2T SM}. \ Gauge fixing of the field theory itself can then be performed
in this 2T field theory and it has been shown (with further evidence in the
present paper) that the resulting lower-dimensional field equations agree with
the wavefunction equations obtained by first quantization of the gauge fixed
system in the worldline formalism \cite{gauge1}\cite{gauge2}. \ 2T field
theory has the advantage that any potential quantum ordering ambiguities of
the first quantized theory are automatically resolved.

More importantly, the field interactions introduced in flat $d+2$ dimensions
are consistent with field interactions in flat or curved $\left(  d-1\right)
+1$ dimensions, but restricts to some extent the possible interactions in the
lower dimension. Interestingly, for the $4+2$ case, the emergent theory in
flat $3+1$ dimensions allows most renormalizable interactions that correspond
to dimension 4 operators. The restriction that emerge on those interactions
are quite interesting:

\begin{itemize}
\item Dimensionful parameters such as masses are not permitted by the gauge
symmetries in the 4+2 theory, so in the emergent $3+1$ theory masses can only
emerge either from the extra dimensions as outlined above, or from spontaneous
breakdown as outlined in \cite{2T SM}. This mass feature may also be helpful
for a resolution of the gauge hierarchy problem with a mechanism that is
different from supersymmetry, but this issue remains to be better understood
in the quantum analysis of the theory.

\item Furthermore \textit{renormalizable} terms of the form $Tr\left(
F_{\mu\nu}F_{\lambda\sigma}\right)  \varepsilon^{\mu\nu\lambda\sigma}$ that
cause the strong CP violation end up having coefficients that are required to
vanish as a gauge choice in the process of reducing from flat 4+2 to flat 3+1
dimensions. This leads to the resolution of the strong CP problem \cite{2T SM}.
\end{itemize}

Given these attractive features of the 4+2 approach, we are then tempted to
propose a new gauge principle in interacting field theory in 3+1 dimensions:
\begin{equation}%
\begin{tabular}
[c]{l}%
The 2T gauge principle - requiring that a theory in 3+1 dimensions\\
be the gauge-fixed version of a 2T theory in 4+2 dimensions. \nonumber
\end{tabular}
\end{equation}
Remarkably, the Standard Model of Particles and Forces (SM) satisfies this
principle \cite{2T SM} as already mentioned. \ While this was demonstrated by
using a specific gauge (namely the "massless particle gauge" in Table 1), the
possibility of choosing other gauges listed in Tables 1,2, implies the
existence of dual versions of the Standard Model in 3+1 dimensions.

Besides the resolution of the strong CP problem, and offering new points of
views on the origin of mass, the 2T gauge principle enunciated above has
further phenomenogical implications for unified theories, SUSY structures, and
cosmology which still need to be worked out (for some comments see \cite{2T
SM}). The duals of the Standard Model or its grand unified and/or
supersymmetric extensions, are likely to suggest additional effects of
phenomenological interest at the LHC.

The current paper is a first step in the investigation of 2T-physics dualities
directly in the field theory formalism. \ For simplicity, in this paper, we
will deal only with the 2T scalar field theory in $d+2$ dimensions. We will
show that a particular class of gauge choices results in a family of
Klein-Gordon (KG) action functionals for the conformal scalar field
propagating on different spacetimes in $\left(  d-1\right)  +1$ dimensions and
interacting locally.

The emergent spacetime metrics in our simplest class of examples use a
generalization of the "relativistic massless particle" gauge to a family of
metrics including the following special cases (even more interesting cases not
in this class appear in Tables 1,2 below)
\begin{equation}%
\begin{tabular}
[c]{l}%
the flat spacetime,\\
the AdS$_{d-n}\times$S$^{n}$ spacetimes,\\
the maximally symmetric spacetimes,\\
the spacetime with a general function $\alpha\left(  x\right)  ,$\\
the Robertson-Walker cosmological spacetimes,\\
the cosmological constant spacetime,\\
the S$^{d-1}\times$R spacetime,\\
the general conformally flat spacetime
\end{tabular}
\ \label{list}%
\end{equation}
It will be shown that interacting KG field theory actions in these curved
spaces have the following properties:

\begin{itemize}
\item Two such emergent field theories, with different background metrics
$g_{\mu\nu}\left(  x\right)  $ and $\tilde{g}_{\mu\nu}\left(  x\right)  $
which are regarded in 1T-physics as different spacetimes, are related to each
other by duality transformations. This is a consequence of the fact that all
the emergent theories are gauge fixed versions of the same theory in $d+2$
dimensions. Remarkably, this duality implies that the theories with
non-trivial spacetimes listed above are all dual to the flat theory with the
Minkowski metric $\eta_{\mu\nu}$.

\item For each fixed background metric $g_{\mu\nu}\left(  x\right)  $ in
$\left(  d-1\right)  +1$ dimensions listed above, the KG field action has a
hidden SO$\left(  d,2\right)  $ global symmetry which is the same as the
original SO$\left(  d,2\right)  $ symmetry of the 2T field theory in $d+2$
dimensions. The explicit form of the SO$\left(  d,2\right)  $ generators,
expressed as transformations of the KG field $\phi\left(  x\right)  $ in the
non-trivial backgrounds $g_{\mu\nu}\left(  x\right)  $ in $\left(  d-1\right)
+1$ dimensions, will also be given.
\end{itemize}

In the rest of the paper, we will first provide in section \ref{gauge choices}
a compendium of gauge choices in the context of the worldline theory. After
that, in Section \ref{2T field} we first review the simplest gauge choice
(massless particle gauge) in the context of field theory for a scalar field.
This is the gauge that leads to the emergent Standard Model in flat $3+1$
spacetime as discussed in \cite{2T SM}. \ Then in Section \ref{main part}, we
will discuss a class of gauge choices that lead to conformally flat spacetimes
in the context of field theory. Our general treatment is specialized to some
interesting cases of emergent spacetimes listed in (\ref{list}) that are often
discussed in the literature in a variety of field theoretic applications. Our
approach shows that the hidden SO$\left(  d,2\right)  $ symmetries in these
field theories and the duality relations among them have often not been
noticed in the past.

The hidden symmetries and dualities apply also to the case of the Standard
Model, as will be shown in a companion paper. . They are verifiable directly
in $\left(  d-1\right)  +1$ dimensions through computation and experiment.
These, together with similar dualities that follow from more interesting gauge
choices listed in Tables 1,2 (and commented on below), are just the tip of an
\textquotedblleft iceberg\textquotedblright\ signaling a unified master theory
in $d+2$ dimensions.

\section{Gauge choices \label{gauge choices}}

In this section, we provide a list of known gauge choices in the worldline
formalism for the simplest 2T-physics systems, which is the free spinless
particle in $d+2$ dimensions subject to the Sp$\left(  2,R\right)  $ gauge
symmetry. These gauge choices have their equivalents in the field theory
formalism, so it is useful to be guided by the worldline theory to understand
their physical meaning in terms of 1T-physics.

The free 2T-physics particle in flat space is described by the action
\[
S=\int d\tau\frac{1}{2}\varepsilon^{ij}\left(  D_{\tau}X_{i}^{M}\right)
X_{j}^{N}\eta_{MN},
\]
where $X_{i}^{M}\left(  \tau\right)  \equiv\left(
\genfrac{}{}{0pt}{}{X^{M}\left(  \tau\right)  }{P_{M}\left(  \tau\right)  }%
\right)  ,$ $i=1,2,$ is phase space considered as doublets under Sp$\left(
2,R\right)  ,$ $\varepsilon^{ij}=\left(
\genfrac{}{}{0pt}{}{0}{-1}%
\genfrac{}{}{0pt}{}{1}{0}%
\right)  ^{ij}$ is the antisymmetric Sp$\left(  2,R\right)  $ metric, and
$D_{\tau}X_{i}^{M}=\partial_{\tau}X_{i}^{M}-A_{i}^{~j}X_{j}^{M}$ is the
Sp$\left(  2,R\right)  $ gauge covariant derivative, with the 3 gauge
potentials $A^{ij}=\varepsilon^{ik}A_{k}^{~j}=\left(
\genfrac{}{}{0pt}{}{A}{C}%
\genfrac{}{}{0pt}{}{C}{B}%
\right)  .$ More explicitly, after dropping a total derivative, we can write
the action in the form
\begin{equation}
S=\int d\tau\left\{  \dot{X}^{M}P_{M}-\frac{1}{2}AX\cdot X-\frac{1}{2}BP\cdot
P-CX\cdot P\right\}  .\label{master}%
\end{equation}
The Sp$\left(  2,R\right)  $ gauge generators are\footnote{More generally, the
generators of Sp$\left(  2,R\right)  $ are more complicated expressions
$Q_{ij}\left(  X,P\right)  $ with $i,j=1,2,$ that depend on the background
fields. All possible background fields can be included \cite{2tbacgrounds}.}
$Q_{ij}=X_{i}\cdot X_{j}=\left(  X\cdot X\right)  ,\left(  P\cdot P\right)
,\left(  X\cdot P\right)  .$ This action has an evident global symmetry
SO$\left(  d,2\right)  $ as the symmetry of the flat metric $\eta_{MN}$ in the
dot product which defines the three Sp$\left(  2,R\right)  $ gauge generators
$X\cdot P=X^{M}P^{N}\eta_{MN},$ etc.. In the gauge fixed versions below the
SO$\left(  d,2\right)  $ symmetry will turn into a hidden global symmetry.

After two Sp$\left(  2,R\right)  $ gauge choices, as given in the tables
below, all components of $X^{M},$ $P^{M}$ are expressed in terms of the
remaining phase space $t,\vec{x},H,\vec{p}$ in $\left(  d-1\right)  +1$
dimensions. Then the action reduces to the form
\begin{equation}
S=\int d\tau\left\{  \overrightarrow{\dot{x}}\cdot\vec{p}-\dot{t}H-\frac{1}%
{2}Bf\left(  t,\vec{x},H,\vec{p}\right)  \right\}
\end{equation}
which describes a particle (not necessarily Lorentz or even rotation
covariant) subject to a generalized $\tau$ reparametrization symmetry. In a
subset of cases the action may take a more covariant form, such as $S=\int
d\tau\left\{  \dot{x}^{\mu}p_{\mu}-\frac{1}{2}Bg^{\mu\nu}\left(  x\right)
p_{\mu}p_{\nu}\right\}  .$ In the gauge $t\left(  \tau\right)  =\tau,$ many
Hamiltonians $H\left(  \vec{x},\vec{p}\right)  $ emerge by solving the
remaining constraint $f\left(  t,\vec{x},H,\vec{p}\right)  =0$.

In the completely gauge fixed version with $\dot{t}=1$ the action becomes
simply
\begin{equation}
S=\int d\tau\left\{  \overrightarrow{\dot{x}}\cdot\vec{p}-H\left(  \vec
{x},\vec{p}\right)  \right\}
\end{equation}
in terms of the unconstrained phase space in 1T-physics. Different
Hamiltonians emerge because time $t$ has been embedded in the 2T phase space
in 4+2 dimensions in different ways, as seen in the explicit parameterizations
$X^{M}\left(  t,\vec{x},H,\vec{p}\right)  ,$ $P^{M}\left(  t,\vec{x},H,\vec
{p}\right)  $ given in Tables 1,2 below. These give a compilation of some
gauge choices that already appeared in previous papers, as well as other gauge
choices that had remained unpublished. The tables are not exhaustive since all
possible gauge choices are not known.

The two tables differ only in the choice of convenient components $M=\left(
+^{\prime},-^{\prime},(m\text{ or }\mu)\right)  $ versus $M=\left(  0^{\prime
},0,1^{\prime},i\right)  $ for parameterizing the gauge choices, where
$X^{\pm^{\prime}}=\frac{1}{\sqrt{2}}\left(  X^{0^{\prime}}\pm X^{1^{\prime}%
}\right)  $. The last column in Table 1 is labeled by $\mu$ in the simple case
$X^{\mu}\left(  x\right)  =x^{\mu}$ and is labeled by $m$ otherwise, where $m$
implies $\left(  \mu\oplus i\right)  $ or more general possibilities. The
total number of dimensions labelled by $m=\mu$ or $m=\left(  \mu\oplus
i\right)  $ or $I=\left(  1^{\prime},i\right)  $ is always $d.$

Explanatory comments on the entries in both tables will be given following
general remarks. For more detailed information on these gauges the reader can
consult \cite{gauge1}-\cite{gauge3}. \
\begin{gather*}%
\begin{tabular}
[c]{||l|l|l|l|l||}\hline\hline
Gauge choice &  & \multicolumn{1}{||l|}{$+^{\prime}$} &
\multicolumn{1}{||l|}{$-^{\prime}$} & \multicolumn{1}{||l||}{$m=\left(
\mu\oplus i\right)  $, $\mu=0,1,\cdots$}\\\hline\hline
$%
\genfrac{}{}{0pt}{}{\text{{\small Relativistic}}{\small \ \;\;\;\;\;\;\;\;}%
}{\text{{\small massless~particle}}}%
$ & $X^{M}=$ & $~1$ & $\frac{1}{2}x^{2}$ & $x^{\mu}$\\\cline{3-5}%
$p^{2}=0$ & $P^{M}=$ & $~0$ & $x\cdot p$ & $p^{\mu}$\\\hline\hline
AdS$_{d-n}\times$S$^{n}${\small \ \ } & $X^{M}=$ & $\frac{R_{{\small 0}}^{2}%
}{\left\vert \overrightarrow{y}\right\vert }$ & $\frac{1}{2\left\vert
\overrightarrow{y}\right\vert }(x^{2}+\overrightarrow{y}^{2})$ &
$\frac{R_{{\small 0}}}{\left\vert \vec{y}\right\vert }x^{\mu},~\frac
{R_{{\small 0}}}{\left\vert \vec{y}\right\vert }\vec{y}^{i}$\\\cline{3-5}%
$\vec{y}^{2}(p^{2}+\vec{k}^{2})=0$ & $P^{M}=$ & $~0$ & $\frac{\left\vert
\overrightarrow{y}\right\vert }{R_{{\small 0}}^{2}}(x\cdot p+\vec{y}\cdot
\vec{k})$ & $\frac{\left\vert \vec{y}\right\vert }{R_{{\small 0}}}p^{\mu
},~\frac{\left\vert \vec{y}\right\vert }{R_{{\small 0}}}\vec{k}^{i}%
$\\\hline\hline
$%
\genfrac{}{}{0pt}{}{\text{{\small Maximally}}{\small ~~~~~~~~~}%
}{\text{{\small Symmetric Spaces}}}%
$ & $X^{M}=$ & ${\small 1+}\sqrt{{\small 1-Kx}^{2}}$ & $\frac{x^{2}/2}%
{1+\sqrt{1-Kx^{2}}}$ & $x^{\mu}$\\\cline{3-5}%
$p^{2}-\frac{K~\left(  x\cdot p\right)  ^{2}}{1-Kx^{2}}=0$ & $P^{M}=$ & $~0$ &
$\frac{\sqrt{1-Kx^{2}}}{1+\sqrt{1-Kx^{2}}}x\cdot p$ & $p^{\mu}-\frac{Kx\cdot
p~x^{\mu}}{1+\sqrt{1-Kx^{2}}}$\\\hline\hline
{\small Free function\ }$\alpha\left(  x\right)  $ & $X^{M}=$ & $x^{2}%
+\alpha\left(  x\right)  $ & $\frac{x^{2}/2}{x^{2}+\alpha\left(  x\right)  }$
& $x^{\mu}$\\\cline{3-5}%
$p^{2}+\frac{4\alpha\left(  x\right)  \left(  x\cdot p\right)  ^{2}}{\left(
x^{2}-\alpha\left(  x\right)  \right)  ^{2}}=0$ & $P^{M}=$ & $~0$ &
$\frac{x\cdot p}{\alpha\left(  x\right)  -x^{2}}$ & $p^{\mu}-\frac{2x\cdot
p~}{x^{2}-\alpha\left(  x\right)  }x^{\mu}$\\\hline\hline
$\underset{g_{\mu\nu}=e_{\mu}^{m}\left(  x\right)  e_{\nu}^{n}\left(
x\right)  \eta_{mn}}{\text{{\small Conformally flat}}}$ & $X^{M}=$ & $\pm
e^{\sigma\left(  x\right)  }$ & $\pm\frac{1}{2}e^{\sigma\left(  x\right)
}q^{2}\left(  x\right)  $ & $\underset{e_{\mu}^{m}\left(  x\right)  \equiv\pm
e^{\sigma\left(  x\right)  }\frac{\partial q^{m}\left(  x\right)  }{\partial
x^{\mu}}}{\pm e^{\sigma\left(  x\right)  }q^{m}\left(  x^{\mu}\right)  }%
$\\\cline{3-5}%
$g^{\mu\nu}\left(  x\right)  p_{\mu}p_{\nu}=0$ & $P^{M}=$ & $~0~$ &
$q^{m}\left(  x\right)  e_{m}^{\mu}\left(  x\right)  p_{\mu}$ & $~e_{m}^{\mu
}\left(  x\right)  p_{\mu}$\\\hline\hline
$%
\genfrac{}{}{0pt}{}{\text{{\small Relativistic}}{\small \ \;\;\;\;\;}%
}{\text{{\small massive~particle}}}%
$ & $X^{M}=$ & $\frac{1+a}{2a}$ & $\frac{x^{2}a}{1+a}$ & $x^{\mu}%
~~~_{a\equiv\sqrt{1+\frac{m^{2}x^{2}}{\left(  x\cdot p\right)  ^{2}}}}%
$\\\cline{3-5}%
$p^{2}+m^{2}=0$ & $P^{M}=$ & $\frac{-m^{2}}{2ax\cdot p}$ & $a~x\cdot p$ &
$p^{\mu}$\\\hline\hline
$%
\genfrac{}{}{0pt}{}{\text{{\small Non-relativistic}}%
}{\text{{\small massive~particle}}}%
$ & $X^{M}=$ & $~t$ & $\frac{\mathbf{r\cdot p}-tH}{m}$ & $X^{0}=\pm\left\vert
\mathbf{r-}\frac{t}{m}\mathbf{p}\right\vert ,~X^{i}=\mathbf{r}^{i}%
$\\\cline{3-5}%
$H-\frac{\mathbf{p}^{2}}{2m}=0$ & $P^{M}=$ & $~m$ & $H$ & $P^{0}%
=0,\;P^{i}=\mathbf{p}^{i}$\\\hline\hline
\end{tabular}
\\
\text{Table1: Parametrization of }X^{M},P^{M}\text{ for }M=\left(  +^{\prime
},-^{\prime},(m\text{ or }\mu)\right)
\end{gather*}

In each gauge choice two degrees of freedom in $\left(  X^{M}\left(
\tau\right)  ,P_{M}\left(  \tau\right)  \right)  $ have been gauge fixed for
all $\tau$ and the two constraints $X^{2}=X\cdot P=0$ have been explicitly
solved to give all components of $X^{M},P_{M}$ in terms of the remaining
degrees of freedom $t,\vec{x},H,\vec{p}$. The third Sp$\left(  2,R\right)  $
constraint $P^{2}=0$ is equivalent to the constraint among the remaining
degrees of freedom $t,\vec{x},H,\vec{p}$ as listed in the first column.

For example, in the case of the massive non-relativistic particle in Table 1,
the two gauge choices are $P^{+^{\prime}}\left(  \tau\right)  =m$ and
$P^{0}\left(  \tau\right)  =0,$ while the solution of $X^{2}=X\cdot P=0$ is
given by $X^{-^{\prime}}=\frac{\vec{r}\cdot\vec{p}-tH}{m}$ and $X^{0}%
=\pm\left\vert \vec{r}\mathbf{-}t\vec{p}/m\right\vert ,$ where $t\left(
\tau\right)  $ is a function of $\tau$ and is canonical to $H\left(
\tau\right)  .$ For the remaining gauge symmetry we can choose $t\left(
\tau\right)  =\tau$ and the remaining constraint gives $0=P^{2}=-2P^{+^{\prime
}}P^{-^{\prime}}-P_{0}^{2}+P_{i}^{2}=-2mH+0+\vec{p}^{2},$ which is solved by
the non-relativistic Hamiltonian $H=\frac{\vec{p}^{2}}{2m}$ listed in the
first column.
\begin{gather*}%
\begin{tabular}
[c]{||l|l|l|l|l||}\hline\hline
Gauge choice & $M$ & \multicolumn{1}{||l|}{$0^{\prime}$} &
\multicolumn{1}{||l|}{$0$} & \multicolumn{1}{||l||}{$I=\left(  1^{\prime
},i\right)  $}\\\hline\hline
$%
\genfrac{}{}{0pt}{}{\text{{\small Robertson-Walker}}{\small \ }r<R_{{\small 0}%
}}{\text{{\small (closed universe)}}}%
$ & $X^{M}=$ & $a\left(  t\right)  \cos(\int^{t}\frac{dt^{\prime}}%
{a(t^{\prime})})$ & $a\left(  t\right)  \sin(\int^{t}\frac{dt^{\prime}%
}{a(t^{\prime})})$ & $%
\genfrac{}{}{0pt}{}{X^{i}=\mathbf{r}^{i}a\left(  t\right)  /R_{{\small 0}%
}~~~~~~}{X^{1^{\prime}}=\pm a\left(  t\right)  \sqrt{1-\frac{r^{2}%
}{R_{{\small 0}}^{2}}}}%
$\\\cline{3-5}%
$%
\genfrac{}{}{0pt}{}{-H^{2}+\frac{R_{{\small 0}}^{2}}{a^{2}\left(  t\right)
}(\mathbf{p}^{2}-\frac{(\mathbf{r\cdot p)}^{2}}{R_{{\small 0}}^{2}})=0}{{}}%
$ & $P^{M}=$ & $-H\sin(\int^{t}\frac{dt^{\prime}}{a(t^{\prime})})$ &
$H\cos(\int^{t}\frac{dt^{\prime}}{a(t^{\prime})})$ & $%
\genfrac{}{}{0pt}{}{P^{i}=\frac{R_{{\small 0}}}{a\left(  t\right)
}(\mathbf{p}^{i}-\frac{\mathbf{r\cdot p}}{R_{{\small 0}}^{2}}\mathbf{r}%
^{i})}{P^{1^{\prime}}=\mp\frac{\mathbf{r\cdot p}}{a\left(  t\right)  }%
\sqrt{1-\frac{r^{2}}{R_{{\small 0}}^{2}}}}%
$\\\cline{2-5}%
$%
\genfrac{}{}{0pt}{}{\text{{\small Robertson-Walker}}{\small \ }%
r>0}{\text{{\small (open universe)}}}%
$ & $X^{M}=$ & $a\left(  t\right)  \sinh(\int^{t}\frac{dt^{\prime}%
}{a(t^{\prime})})$ & ($\pm)^{\prime}a\left(  t\right)  \sqrt{1+\frac{r^{2}%
}{R_{{\small 0}}^{2}}}$ & $%
\genfrac{}{}{0pt}{}{X^{i}=\mathbf{r}^{i}a\left(  t\right)  /R_{{\small 0}%
}~~~~~~~~}{X^{1^{\prime}}=\pm a\left(  t\right)  \cosh(\int^{t}\frac
{dt^{\prime}}{a(t^{\prime})})}%
$\\\cline{3-5}\cline{3-5}%
$%
\genfrac{}{}{0pt}{}{-H^{2}+\frac{R_{{\small 0}}^{2}}{a^{2}\left(  t\right)
}(\mathbf{p}^{2}+\frac{(\mathbf{r\cdot p)}^{2}}{R_{{\small 0}}^{2}})=0}{{}}%
$ & $P^{M}=$ & $\pm H\cosh(\int^{t}\frac{dt^{\prime}}{a(t^{\prime})})$ &
($\pm)^{\prime}\frac{\mathbf{r\cdot p}}{a\left(  t\right)  }\sqrt
{1+\frac{r^{2}}{R_{{\small 0}}^{2}}}$ & $%
\genfrac{}{}{0pt}{}{P^{i}=\frac{R_{{\small 0}}}{a\left(  t\right)
}(\mathbf{p}^{i}+\frac{\mathbf{r\cdot p}}{R_{{\small 0}}^{2}}\mathbf{r}%
^{i})}{P^{1^{\prime}}=H\sinh(\int^{t}\frac{dt^{\prime}}{a(t^{\prime})})}%
$\\\hline\hline
$\underset{\Lambda\equiv{\small ~}\frac{3}{R_{{\small 0}}^{2}}~>~0}%
{\text{{\small Cosmological\ constant}}}$ & $X^{M}=$ & $\sqrt{R_{{\small 0}%
}^{2}-r^{2}}\sinh\frac{t}{R_{{\small 0}}}$ & $~~R_{{\small 0}}\;\;$ & $%
\genfrac{}{}{0pt}{}{X^{i}=\mathbf{r}^{i}\;\;\;\;\;\;\;\;\;\;}{X^{1^{\prime}%
}=\pm\sqrt{R_{{\small 0}}^{2}-r^{2}}\cosh\frac{t}{R_{{\small 0}}}}%
$\\\cline{3-5}%
$%
\genfrac{}{}{0pt}{}{-H^{2}(1-\frac{r^{2}}{R_{{\small 0}}^{2}})+(\mathbf{p}%
^{2}+\frac{(\mathbf{r\cdot p)}^{2}}{R_{{\small 0}}^{2}-r^{2}})=0}{{}}%
$ & $P^{M}=$ & $\pm\frac{H}{R_{{\small 0}}}\sqrt{R_{{\small 0}}^{2}-r^{2}%
}\cosh\frac{t}{R_{{\small 0}}}$ & $~\frac{R_{{\small 0}}\mathbf{r\cdot p}%
}{R_{{\small 0}}^{2}-r^{2}}$ & $%
\genfrac{}{}{0pt}{}{P^{i}=\mathbf{p}^{i}+\frac{\mathbf{r\cdot p}%
}{R_{{\small 0}}^{2}-r^{2}}\mathbf{r}^{i},~}{P^{1^{\prime}}=\frac
{H}{R_{{\small 0}}}\sqrt{R_{{\small 0}}^{2}-r^{2}}\sinh\frac{t}{R_{{\small 0}%
}}}%
$\\\cline{2-5}%
$\underset{\Lambda\equiv{\small ~-}\frac{3}{R_{{\small 0}}^{2}}~<~0}%
{\text{{\small Cosmological\ constant}}}$ & $X^{M}=$ & $\sqrt{R_{{\small 0}%
}^{2}+r^{2}}\sin\frac{t}{R_{{\small 0}}}$ & $\mp\sqrt{R_{{\small 0}}^{2}%
+r^{2}}\cos\frac{t}{R_{{\small 0}}}$ & $%
\genfrac{}{}{0pt}{}{X^{i}=\overrightarrow{r}^{i}\;}{X^{1^{\prime}%
}=R_{{\small 0}}~}%
$\\\cline{3-5}%
$%
\genfrac{}{}{0pt}{}{-H^{2}(1+\frac{r^{2}}{R_{{\small 0}}^{2}})+(\mathbf{p}%
^{2}-\frac{(\mathbf{r\cdot p)}^{2}}{R_{{\small 0}}^{2}+r^{2}})=0}{{}}%
$ & $P^{M}=$ & $\pm\frac{H}{R_{{\small 0}}}\sqrt{R_{{\small 0}}^{2}+r^{2}}%
\cos\frac{t}{R_{{\small 0}}}$ & $\frac{H}{R_{{\small 0}}}\sqrt{R_{{\small 0}%
}^{2}+r^{2}}\sin\frac{t}{R_{{\small 0}}}$ & $%
\genfrac{}{}{0pt}{}{P^{i}=\mathbf{p}^{i}-\frac{\mathbf{r\cdot p}%
}{R_{{\small 0}}^{2}+r^{2}}\mathbf{r}^{i},~}{P^{1^{\prime}}=-\frac
{R_{{\small 0}}\mathbf{r\cdot p}}{R_{{\small 0}}^{2}+r^{2}}}%
$\\\hline\hline
{\small (d--1)-sphere}$\times${\small time} & $X^{M}=$ & $R_{{\small 0}}%
\cos\frac{t}{R_{{\small 0}}}$ & $R_{{\small 0}}\sin\frac{t}{R_{{\small 0}}}$ &
$R_{{\small 0}}\widehat{n}^{I}=%
\genfrac{}{}{0pt}{}{X^{i}=\mathbf{r}^{i}\;\;\;\;~~~~}{X^{1^{\prime}}=\pm
\sqrt{R_{{\small 0}}^{2}-r^{2}}}%
$\\\cline{3-5}%
$%
\genfrac{}{}{0pt}{}{-H^{2}+(\mathbf{p}^{2}+\frac{(\mathbf{r\cdot p)}^{2}%
}{R_{{\small 0}}^{2}-r^{2}})=0}{{}}%
$ & $P^{M}=$ & $-H\sin\frac{t}{R_{{\small 0}}}$ & $H\cos\frac{t}%
{R_{{\small 0}}}$ & $%
\genfrac{}{}{0pt}{}{P^{i}=\mathbf{p}^{i}\;\;\;\;\;}{P^{1^{\prime}}=\mp
\frac{\mathbf{r\cdot p}}{\sqrt{R_{{\small 0}}^{2}-r^{2}}}}%
$\\\hline\hline
H-atom,~$H<0$ & $X^{M}=$ & $\underset{u({\small t)}\equiv\frac{\sqrt{-2mH}%
}{m\alpha}\left(  \mathbf{r\cdot p}-2mHt\right)  }{r\cos u}$ & $r\sin u$ & $%
\genfrac{}{}{0pt}{}{X^{i}=\mathbf{r}^{i}-\frac{r}{m\alpha}\mathbf{r\cdot
p~p}^{i}}{X^{1^{\prime}}=-\frac{r}{m\alpha}\sqrt{-2mH}\mathbf{r\cdot p}}%
$\\\cline{3-5}%
$H=\frac{\mathbf{p}^{2}}{2m}-\frac{\alpha}{r}$ & $P^{M}=$ & $-\frac{m\alpha
}{r\sqrt{-2mH}}\sin u$ & $\frac{m\alpha}{r\sqrt{-2mH}}\cos u$ & $%
\genfrac{}{}{0pt}{}{P^{i}=\mathbf{p}^{i}~~~~~~~~~~~}{P^{1^{\prime}}=\frac
{1}{\sqrt{-2mH}}\left(  \frac{m\alpha}{r}-\mathbf{p}^{2}\right)  }%
$\\\cline{2-5}%
H-atom,\ $H>0$ & $X^{M}=$ & $\underset{u({\small t)}\equiv\frac{\sqrt{2mH}%
}{m\alpha}\left(  \mathbf{r\cdot p}-2mHt\right)  }{r\cosh u}$ & $\frac
{r}{m\alpha}\sqrt{2mH}\mathbf{r\cdot p}$ & $%
\genfrac{}{}{0pt}{}{X^{i}=\mathbf{r}^{i}-\frac{r}{m\alpha}\mathbf{r\cdot
p~p}^{i}}{X^{1^{\prime}}=r\sinh u}%
$\\\cline{3-5}
& $P^{M}=$ & $\frac{m\alpha}{r\sqrt{2mH}}\sinh u$ & $\frac{1}{\sqrt{2mH}%
}\left(  \frac{m\alpha}{r}-\mathbf{p}^{2}\right)  $ & $%
\genfrac{}{}{0pt}{}{P^{i}=\mathbf{p}^{i}~~~~~~~~~~~}{P^{1^{\prime}}%
=\frac{m\alpha}{r\sqrt{2mH}}\cosh u}%
$\\\hline\hline
\end{tabular}
\\
\text{Table2 : Parametrization of }X^{M},P^{M}~\text{for }M=\left(  0^{\prime
},0,I\right)
\end{gather*}

The different Hamiltonians $H\left(  \vec{x},\vec{p}\right)  $ that emerge
from such gauge choices are all holographic representatives of the same master
theory in Eq.(\ref{master}) and therefore must form a set of dual theories
that are transformed into each other by the original gauge symmetry Sp$\left(
2,R\right)  .$ Furthermore the emergent actions all must have hidden global
symmetry SO$\left(  d,2\right)  .$ The generators of SO$\left(  d,2\right)  $
are $L^{MN}=\varepsilon^{ij}X_{i}^{M}X_{j}^{N}=X^{M}P^{N}-X^{M}P^{N},$ which
are evidently gauge invariant under Sp$\left(  2,R\right)  $, and are
conserved $\partial_{\tau}L^{MN}=0$ by using the original equations of motion
that follow from (\ref{master}). In the respective phase spaces these take the
following non-linear forms
\begin{equation}
L^{MN}=X^{M}\left(  t,\vec{x},H,\vec{p}\right)  ~P^{N}\left(  t,\vec{x}%
,H,\vec{p}\right)  -X^{N}\left(  t,\vec{x},H,\vec{p}\right)  ~P^{M}\left(
t,\vec{x},H,\vec{p}\right)  .
\end{equation}
Under Poisson brackets in the respective phase spaces these $L^{MN}$ close to
form the Lie algebra of SO$\left(  d,2\right)  .$ Here $\left(  t,H\right)  $
can be treated as canonical conjugates. But, it is also possible to make the
final gauge choice $t\left(  \tau\right)  =\tau;$ in that case $\tau$ is
treated as a constant and $H$ is replaced by the solution of the last
constraint $f\left(  t,\vec{x},H,\vec{p}\right)  =0,$ which gives $H=H\left(
\vec{x},\vec{p}\right)  $ as dependent on the remaining canonical variables.

In the quantum theory, canonical conjugates must be ordered to insure that the
$L^{MN}$ satisfy the SO$\left(  d,2\right)  $ Lie algebra. This was done
successfuly for some of the examples in Tables 1,2 in the first quantization
approach \cite{gauge1}\cite{gauge2}. In the field theory context discussed in
this paper, the quantum ordering is automatically achieved, as discussed in
section \ref{LMN gen}.

The $L^{MN}$ are constants of motion and they have the same gauge invariant
value in each phase space. So any function of the $L^{MN}$ has identical
values in any of the phase spaces that appear in Tables 1,2. Therefore the
$L^{MN}$ are the key to the dualities among these systems. The duality
transformation is the original Sp$\left(  2,R\right)  ,$ which transforms one
phase space (a given gauge choice) to another. Through the gauge invariant (or
duality invariant) $L^{MN}$ one can establish an infinite set of duality
relations among these systems. These can be checked through computation or
through experiment. In the first quantized theory, one consequence of this
duality is that all of the systems in Tables 1,2 provide Hilbert spaces that
must span the same representation of SO$\left(  d,2\right)  $ (but in
different bases defined by diagonalizing the Hamiltonian $H$). The SO$\left(
d,2\right)  $ Casimir eigenvalues $C_{n}$ of this universal representation are
independent of the gauge choice; for example the quadratic Casimir eigenvalue
at the quantum level is given by $C_{2}=\frac{1}{2}L^{MN}L_{MN}=1-d^{2}/4$
\cite{2T basics}. The universal value of the Casimirs is one of the
consequences of duality that is independent of the details of a particular
quantum state, and can be verified easily to be true for each physical system
described in the Tables.

Similarly all the other Casimirs operators are fixed\footnote{In the classical
theory $C_{2}=0,$ and similarly all $C_{n}=0,$ follow from the constraints
$X^{2}=P^{2}=X\cdot P=0.$ But in the quantum theory, ordering of operators
lead to the non-zero eigenvalues of $C_{n}$ that correspond to the singleton
representation.}, and the resulting unitary representation is the singleton of
SO$\left(  d,2\right)  .$ In the same singleton space, each system in Tables
1,2 corresponds to a different choice of basis labelled by the eigenvalues of
simultaneous observables, one of which is the choice of Hamiltonian $H$
(choice of time) in the respective phase space. The transformation $\langle
basis~1|~basis~2\rangle$ from one complete basis to another within the same
singleton representation should be understood as the duality transformation at
the quantum level.

The following comments give further information on the entries in Table 1.

\begin{enumerate}
\item For the maximally symmetric space the canonical pairs are $\left(
x^{\mu},p_{\mu}\right)  .$ The curved space on which the particle propagates
is determined by the flat SO$\left(  d,2\right)  $ metric $ds^{2}=dX^{M}%
dX^{N}\eta_{MN}=-2dX^{+^{\prime}}dX^{-^{\prime}}+dX^{\mu}dX^{\nu}\eta_{\mu\nu
}$ where $\eta_{\mu\nu}$ is the Minkowski metric in $\left(  d-1\right)  +1$
dimensions. This metric becomes $ds^{2}=dx^{\mu}dx^{\nu}g_{\mu\nu}\left(
x\right)  ,$ with $g_{\mu\nu}=\eta_{\mu\nu}+\frac{K}{1-Kx^{2}}x_{\mu}x_{\nu}.$
The constraint listed in the first column $P\cdot P=p^{2}-\frac{K~\left(
x\cdot p\right)  ^{2}}{1-Kx^{2}}=g^{\mu\nu}\left(  x\right)  p_{\mu}p_{\nu}=0$
involves the inverse of the metric. When the canonical operators are properly
ordered in the quantum theory, this constraint becomes the Laplacian in curved
space for the conformal scalar (including the curvature term) given in
Eq.(\ref{confflat}), as we will show\footnote{The same result would follow in
the first quantization approach by ordering properly the $L^{MN}\left(
t,\vec{x},H,\vec{p}\right)  $ and insuring that the quantum constraint listed
in the first column is invariant (or properly transforms) under it. This was
the method used in \cite{gauge2} as demonstrated for the AdS$_{d-n}\times
$S$^{n}$ case. In the present paper, the field theory approach automatically
takes care of all ordering issues.} in section \ref{main part}. The Riemann
scalar curvature for this space is a constant $R=K$ , and evidently it reduces
to flat space (first entry in Table 1) if the curvature modulus $K$ in the
gauge choice vanishes $K\rightarrow0.$ Some well known maximally symmetric
spaces include DeSitter space with $K>0$ and Anti-deSitter space with $K<0.$
For more information on maximally symmetric spaces see \cite{Weinberg}.

\item For AdS$_{d-n}\times$S$^{n},$ the canonical pairs are the $\left(
x^{\mu}\left(  \tau\right)  ,p_{\mu}\left(  \tau\right)  \right)  $ in $d-n-1$
dimensions and the $\left(  y^{i}\left(  \tau\right)  ,k_{i}\left(
\tau\right)  \right)  $ in $n+1$ dimensions, for a total of $d$ dimensions.
The curved space on which the particle propagates is given by $ds^{2}%
=dX^{M}dX^{N}\eta_{MN}=\frac{R_{{\small 0}}^{2}}{y^{2}}\left[  \left(
dx^{\mu}\right)  ^{2}+\left(  d\overrightarrow{y}^{i}\right)  ^{2}\right]
=\frac{R_{{\small 0}}^{2}}{y^{2}}\left(  \left(  dx\right)  ^{2}+\left(
dy\right)  ^{2}\right)  +R_{{\small 0}}^{2}\left(  d\Omega_{n}\right)  ^{2}.$
In the last expression $y\equiv\left\vert \vec{y}\right\vert $ and then
$\frac{R_{{\small 0}}^{2}}{y^{2}}\left(  \left(  dx\right)  ^{2}+\left(
dy\right)  ^{2}\right)  $ is the AdS$_{d-n}$ metric, while $\left(
d\Omega_{n}\right)  ^{2}$ is the $S^{n}$ metric build from the unit vector
$\vec{y}/y$ embedded $n+1$ dimensions. Note that our construction in terms of
the SO$\left(  d,2\right)  $ vector $X^{M}$ shows that the full symmetry of
the spacetime AdS$_{d-n}\times$S$^{n}$ is SO$\left(  d,2\right)  $ and not
only SO$\left(  d-n-1,2\right)  \times$SO$\left(  n+1\right)  $ as it is it
often mentioned in the literature (see \cite{gauge2} for more details).

\item The metric that emerges in the case of the free function $\alpha\left(
x\right)  $ is of the form: $g_{\mu\nu}=\eta_{\mu\nu}-\frac{4\alpha\left(
x\right)  }{\left(  x^{2}+\alpha\left(  x\right)  \right)  ^{2}}x_{\mu}x_{\nu
},$ where $\alpha\left(  x\right)  $ is allowed to be any function of $x^{\mu
}.$ If we specialize to the case of a constant $\alpha$ we see that the space
is asymptotically flat and we obtain simple expressions for its curvature
tensors%
\begin{align}
R_{\lambda\sigma\mu\nu}  &  =\frac{-4\alpha}{\left(  x^{2}-\alpha\right)
^{2}}\eta_{\lambda\lbrack\mu}\eta_{\nu]\sigma}-\frac{8\alpha}{\left(
x^{2}-\alpha\right)  ^{2}\left(  x^{2}+\alpha\right)  }x_{[\mu}\eta
_{\nu][\lambda}x_{\sigma]}~~\text{(if }\alpha=\text{constant )}\\
R_{\lambda\mu}  &  =\eta_{\lambda\mu}\left(  4\alpha\frac{\left(  d-1\right)
\alpha-x^{2}\left(  d-3\right)  }{\left(  x^{2}-\alpha\right)  ^{3}}\right)
+x_{\lambda}x_{\mu}\left(  8\alpha\frac{x^{2}\left(  d-2\right)  -d\alpha
}{\left(  x^{2}-\alpha\right)  ^{3}\left(  x^{2}+\alpha\right)  }\right) \\
R  &  =\frac{-4\alpha}{\left(  x^{2}-\alpha\right)  ^{2}}\left(  d^{2}%
+d\frac{x^{2}+\alpha}{x^{2}-\alpha}-\frac{2x^{2}\left(  x^{2}+\alpha\right)
^{2}}{\left(  x^{2}-\alpha\right)  ^{3}}\right)
\end{align}
$\allowbreak$There are curvature singularities at $x^{2}=\alpha,$ and there
are also values of $x^{2}$ where the curvature scalar vanishes. If
$\alpha\left(  x\right)  $ is allowed to be a function rather than a constant
then the expressions for the curvature tensors are more involved, but
generically we expect a space with curvature singularities and zeroes.

\item The conformally flat case is the most general gauge in which all
components of position $X^{M}\left(  x\right)  $ in $d+2$ dimensions are a
functions of only position $x^{\mu}$ (and not momentum $p_{\mu}$) in $\left(
d-1\right)  +1$ dimensions. The main part of this paper starting with the next
section will be involved with the equivalent of this gauge choice in the
context of 2T field theory. \ The conformally flat case is a generalized
version of the four items that precedes it in Table-1. $q^{m}\left(  x\right)
$ and $\sigma\left(  x\right)  $ can be chosen arbitrarily as functions of the
spacetime coordinates $x^{\mu}\left(  \tau\right)  $, while the canonical
conjugates are $\left(  x^{\mu}\left(  \tau\right)  ,p_{\mu}\left(
\tau\right)  \right)  $. The curved space on which the particle propagates is
given by $ds^{2}=dX^{M}dX^{N}\eta_{MN}=dx^{\mu}dx^{\nu}g_{\mu\nu}\left(
x\right)  $ with
\begin{equation}
g_{\mu\nu}\left(  x\right)  =e_{\mu}^{m}\left(  x\right)  e_{\nu}^{n}\left(
x\right)  \eta_{mn},\;\;e_{\mu}^{m}\left(  x\right)  =e^{\sigma\left(
x\right)  }\frac{\partial q^{m}\left(  x\right)  }{\partial x^{\mu}},
\label{confflat}%
\end{equation}
where $\eta_{mn}$ is the flat Minkowski metric in $d$ dimensions$.$ This is
the general conformally flat metric. Evidently this general parametrization
reproduces the maximally symmetric space by taking $e^{\sigma\left(  x\right)
}q^{m}\left(  x\right)  =\delta_{\mu}^{m}x^{\mu}$ and $e^{\sigma\left(
x\right)  }={\small 1+}\sqrt{{\small 1-Kx}^{2}}.$ It also reproduces the
AdS$_{d-n}\times$S$^{n}$ and the free-function-$\alpha$ cases by taking
$q^{m}\left(  x\right)  =\frac{1}{R_{{\small 0}}}\left(  x^{\mu}%
,\overrightarrow{y}^{i}\right)  $ with $e^{\sigma\left(  x\right)  }%
=\frac{R_{{\small 0}}^{2}}{\left\vert \overrightarrow{y}\right\vert }$ for
AdS$_{d-n}\times$S$^{n}$, and $e^{\sigma\left(  x\right)  }q^{m}\left(
x\right)  =\delta_{\mu}^{m}x^{\mu}$ with $e^{\sigma\left(  x\right)  }%
=\frac{x^{2}}{x^{2}+\alpha\left(  x\right)  }$ for free-function-$\alpha$. The
curvature tensors of the general conformally flat space (\ref{confflat}) are
computed in the Appendix. In particular the scalar curvature is given by
$R=d\left(  1-d\right)  \left[  g^{\mu\nu}\left(  \partial_{\mu}\sigma
\partial_{\nu}\sigma+\frac{2}{d}\partial_{\mu}\partial_{\nu}\sigma\right)
+\frac{2}{d}e^{\nu m}\left(  \partial_{\nu}e_{m}^{\mu}\right)  \partial_{\mu
}\sigma\right]  $ as in Eq.(\ref{R}). \

\item The massive relativistic and non-relativistic gauges are distinguished
from the others in Table-1 by the fact that the expressions for some of the
positions $X^{M}\left(  x,p\right)  $ involve momenta $p$. This does not
happen with the other gauges in which the positions $X^{M}\left(  x\right)  $
are functions of only positions $x.$ In the latter case $X^{M}\left(
x\right)  $ we always get a particle moving on some curved space as explained
above. However, when positions and momenta are mixed in the gauge choices
$X^{M}\left(  x,p\right)  $, the emerging dynamics is more intricate and
cannot be described as due to curved space only. We will make more comments on
this feature at the end of this section.
\end{enumerate}

The following comments give further information on the entries in Table 2. As
in Table 1, the emergent spacetimes in $\left(  d-1\right)  +1$ dimensions for
the cases $X^{M}\left(  x\right)  $ are all described by $ds^{2}=dX^{M}%
dX^{N}\eta_{MN}=-\left(  dX^{0^{\prime}}\right)  ^{2}-\left(  dX^{0}\right)
^{2}+\left(  dX^{1^{\prime}}\right)  ^{2}+\left(  d\vec{X}\right)  ^{2}%
=g_{\mu\nu}\left(  x\right)  dx^{\mu}dx^{\mu},$ which shows that they all are
symmetric under the hidden SO$\left(  d,2\right)  $ global symmetry.

\begin{enumerate}
\item In the Robertson-Walker case the metric is given by
\begin{equation}
ds^{2}=-dt^{2}+a^{2}\left(  t\right)  \left[  \left(  1\mp\frac{r^{2}%
}{R_{{\small 0}}^{2}}\right)  ^{-1}dr^{2}+r^{2}\left(  d\Omega_{d-2}\right)
^{2}\right]  \;%
\genfrac{}{}{0pt}{}{\left(  -\right)  ~r<R_{{\small 0}}}{\left(  +\right)
~r>0}%
\end{equation}
In a cosmological context this, and its more specialized Friedman universe
version, is the spacetime that describes the evolution of the universe as a
whole. \ The \ $\left(  -\right)  $ and $\left(  +\right)  $ cases correspond
to closed and open universes respectively.

\item The gauge labelled as the cosmological constant describes a particle
moving in free space except for the influence of a cosmological constant
$\Lambda$. The metric in this case is given by
\begin{equation}
ds^{2}=-\left(  1-\frac{\Lambda}{3}r^{2}\right)  dt^{2}+\left(  1-\frac
{\Lambda}{3}r^{2}\right)  ^{-1}dr^{2}+r^{2}\left(  d\Omega_{d-2}\right)  ^{2}%
\end{equation}
for either positive or negative $\Lambda=\pm\frac{3}{R_{{\small 0}}^{2}}.$
This a particular form of the deSitter ($\Lambda>0)$ or anti-deSitter
($\Lambda<0$) metric.

\item In the case of the $\left(  d-1\right)  $ sphere$\times$time, the
particle moves on a spacetime described by the metric
\begin{equation}
ds^{2}=dX^{M}dX^{N}\eta_{MN}=-dt^{2}+R_{{\small 0}}^{2}\left(  d\Omega
_{d-1}\right)  ^{2},
\end{equation}
The metric on the sphere $R_{0}^{2}\left(  d\Omega_{d-1}\right)  ^{2}$ is
built from the unit vector $n^{I}$ in $d$ dimensions. If $n^{I}\left(  \vec
{r}\right)  $ is parameterized as given in the table, the metric takes the
form $ds^{2}=-dt^{2}+\left(  1-\frac{r^{2}}{R_{{\small 0}}^{2}}\right)
^{-1}dr^{2}+r^{2}\left(  d\Omega_{d-2}\right)  ^{2}$ for $r<R_{{\small 0}}.$

\item The H-atom gauge is fairly complex since it involves the type of
parametrization $X^{M}\left(  x,p\right)  $ which includes both position and
momentum\footnote{The $\left(  X^{M},P^{M}\right)  $ in our table is related
to the $\left(  \tilde{X}^{M},\tilde{P}^{M}\right)  $ for the H-atom given in
a previous publication \cite{gauge1}, by the relation $X^{M}=r\tilde{X}^{M}$
and $P^{M}=\frac{1}{r}\tilde{P}^{M}.$ This is an Sp$\left(  2,R\right)  $
transformation that does not change the meaning of time or Hamiltonian.
Furthermore, compared to \cite{gauge1} we have replaced $\alpha$ by $m\alpha
.$}. It is worth noting that the vector $X^{I}/r=(X^{1^{\prime}},\vec{X})/r$
as parameterized on the table in the case of $H<0$ is a unit vector $\left(
X^{I}/r\right)  ^{2}=1$ embedded in $d$ dimensions. Similarly the momentum
$rP^{I}=r(P^{1^{\prime}},\vec{P})$ is also a unit vector. For $d=4,$ this
explains the SO$\left(  4\right)  $ symmetry of the H-atom (or planetary
system) Hamiltonian as being due to rotation symmetry in four space
dimensions. Including all the $d+2$ coordinates, one learns that the
non-relativistic \textit{action} that describes the H-atom (i.e. particle in
$1/r$ potential) has the hidden symmetry SO$\left(  d,2\right)  ,$ as do all
the other sytems listed in Tables 1,2.

\item Additional gauge choices of the type $X^{M}\left(  x,p\right)  $ which
includes both position and momentum are easy to generate from the ones listed
above by interchanging the roles of position/momentum for some of the entries
in the process of choosing gauges.
\end{enumerate}

Our focus in this paper is field theory. The first goal is to find the analogs
of gauge choices displayed above in the language of field theory, and use them
to derive many 1T-physics field theories from the same 2T-physics field
theory. This will establish a set of dual field theories which could be used
as a technical tool for computations, as well as for the purposes of
unification leading to a deeper understanding of Nature.

We will see that for the gauge fixing of the 2T field theory we will also need
the curvature scalar for the metrics that appeared in Tables 1-2. Therefore we
collect here the Ricci scalar for these metrics. The Ricci scalar for the
conformally flat scalar in the last item is computed in the Appendix.%
\begin{gather}%
\begin{tabular}
[c]{|l|l|l|}\hline
& metric & curvature scalar $R$\\\hline
{\small Flat space} & $ds^{2}=dx\cdot dx\equiv\eta_{\mu\nu}dx^{\mu}dx^{\nu}$ &
$0$\\\hline
{\small AdS}$_{d-n}\times${\small S}$^{n}$ & $ds^{2}=\frac{R_{{\small 0}}^{2}%
}{y^{2}}(dx\cdot dx+dy^{2})+R_{{\small 0}}^{2}(d\Omega_{n})^{2}$ & $\frac
{1}{R_{{\small 0}}^{2}}\left(  2n-d\right)  \left(  d-1\right)  $\\\hline
{\small Maximally symm.} & $ds^{2}=dx\cdot dx+\frac{K}{1-Kx^{2}}\left(  x\cdot
dx\right)  ^{2}$ & $K$\\\hline
{\small Robertson-Walker }$a\left(  t\right)  $ & $ds^{2}=-dt^{2}%
+a^{2}\left\{
\begin{array}
[c]{c}%
\frac{1}{1\mp r^{2}/R_{{\small 0}}^{2}}dr^{2}\\
+r^{2}\left(  d\Omega_{d-2}\right)  ^{2}%
\end{array}
\right\}  $ & $\left(  {\small d-2}\right)  \left(  {\small d-1}\right)
\left\{
\begin{array}
[c]{c}%
\frac{\ddot{a}}{a}+\left(  \frac{\dot{a}}{a}\right)  ^{2}\\
\pm\frac{1}{a^{2}R_{{\small 0}}^{2}}%
\end{array}
\right\}  $\\\hline
{\small Cosmological const.} & $ds^{2}=\left\{
\begin{array}
[c]{c}%
-\left(  1-\frac{\Lambda}{3}r^{2}\right)  dt^{2}\\
+\frac{dr^{2}}{1-\frac{\Lambda}{3}r^{2}}+r^{2}\left(  d\Omega_{d-2}\right)
^{2}%
\end{array}
\right\}  $ & $d\Lambda$\\\hline
$\left(  {\small d-1}\right)  ${\small -sphere}$\times${\small time} &
$ds^{2}=-dt^{2}+R_{{\small 0}}^{2}\left(  d\Omega_{d-1}\right)  ^{2}$ &
$\frac{1}{R_{{\small 0}}^{2}}\left(  d-2\right)  \left(  d-1\right)  $\\\hline
{\small Free function }$\alpha\left(  x\right)  $ & $ds^{2}=dx\cdot
dx+\frac{4\alpha\left(  x\right)  }{\left(  x^{2}+\alpha\left(  x\right)
\right)  ^{2}}\left(  x\cdot dx\right)  ^{2}$ & $\underset{\text{for }%
\alpha=\text{constant}}{\frac{-4\alpha}{\left(  x^{2}-\alpha\right)  ^{2}%
}(d^{2}+d\frac{x^{2}+\alpha}{x^{2}-\alpha}-\frac{2x^{2}\left(  x^{2}%
+\alpha\right)  ^{2}}{\left(  x^{2}-\alpha\right)  ^{3}})}$\\\hline
{\small Conformally flat} & $ds^{2}=e^{2\sigma\left(  x\right)  }\frac
{q^{m}\left(  x\right)  }{\partial x^{\mu}}\frac{q_{m}\left(  x\right)
}{\partial x^{\nu}}dx^{\mu}dx^{\nu}$ & $\left(  1-d\right)  \left\{
\begin{array}
[c]{c}%
dg^{\mu\nu}\partial_{\mu}\sigma\partial_{\nu}\sigma\\
+2g^{\mu\nu}\left(  \partial_{\mu}\partial_{\nu}\sigma\right) \\
+2e^{\nu m}\partial_{\nu}e_{m}^{\mu}\partial_{\mu}\sigma
\end{array}
\right\}  $\\\hline
\end{tabular}
\nonumber\\
\text{Table 3 - Curvature scalar for metrics in Tables 1,2.} \label{table3}%
\end{gather}

As we will show, all the cases in Tables 1,2 which do not involve momenta in
the gauge choices of $X^{M}\left(  x\right)  $ are easily reproduced in 2T
field theory. For this reason in this first paper we concentrate on these
cases in 2T field theory as presented in Sections \ref{2T field}%
\ref{flat field} and \ref{main part}.\ \ The cases that mix position and
momentum in the gauge choices for $X^{M}\left(  x,p\right)  ,$ such as the
examples of the massive particles, including relativistic, non-relativistic or
H-atom gauges, are harder because the $d+2$ field theory is formulated by
making a distinction between position and momentum $X^{M}$ and $P_{M}%
\rightarrow-i\partial_{M}$ at the outset (see \cite{2T field paper} for some
discussion of the non-relativistic case in field theory). Since these cases
involve mass, which emerges as a modulus from the higher dimensions, they seem
rather interesting as a notion that could relate to the origin of mass. We
plan to discuss this issue in a future paper.

\section{d+2 field theory\label{2T field}}

2T field theory has been fully formulated for fields of spins $0,\frac{1}%
{2},1$ \cite{2T SM}, and to a lesser extent for spin 2 \cite{2T field paper}
and beyond \cite{2tbacgrounds}, and has also been supersymmetrized
\cite{susy2tN1}. The scalar field provides a first example in the present
paper for multiple gauge fixing. Its 2T action is
\begin{equation}
S\left(  \Phi\right)  =Z\int\left(  d^{d+2}X\right)  \delta\left(
X^{2}\right)  \left[  \frac{1}{2}\Phi\partial^{2}\Phi-\gamma\frac{d-2}{2d}%
\Phi^{\frac{2d}{d-2}}\right]  , \label{scalar field action}%
\end{equation}
Here $Z$ is an overall normalization constant that will be determined.

The action was obtained through a BRST procedure \cite{2tbrst2006} consistent
with the underlying Sp$\left(  2,R\right)  $ gauge symmetry of the worldline
theory (\ref{master}). The equations of motion Eqs.(\ref{scalar field eom})
that follow from this action impose the Sp$\left(  2,R\right)  $ gauge singlet
conditions, $X^{2}=X\cdot P=P^{2}=0,$ in the free field case. These free field
equations are equivalent to covariant first quantization of the worldline
theory. The BRST approach of \cite{2tbrst2006} allows interactions of the
special form $V\left(  \Phi\right)  =\gamma\frac{d-2}{2d}\Phi^{\frac{2d}{d-2}%
}$, which is the renormalizable $\Phi^{4}$ interaction for $d+2=4+2$.

Note, in particular the presence of the delta function in the action, which
imposes the $X^{2}=0$ condition, and which is crucial to obtaining three
($X^{2}=0,$ kinematic, and dynamic) 2T equations of motion
Eqs.(\ref{scalar field eom}) from this action with a single field \cite{2T
SM}. \ It should also be emphasized that, due to this $\delta\left(
X^{2}\right)  ,$ the action is not invariant under translations of $X^{M}$.
\ However, one should realize that, for some gauge choices, the translations
in the lower dimensions are included in the original SO$\left(  d,2\right)  $
symmetry. Indeed, in the next section we will describe the emergent field
theory in flat spacetime in the $\left(  d-1\right)  +1$ dimensions that is
obtained by considering the massless particle gauge of Table 1. The SO$\left(
d,2\right)  $ symmetry will then be interpreted as conformal symmetry, which
includes the Poincar\'{e} symmetry with generators $L^{\mu\nu}=x^{\mu}p^{\nu
}-x^{\nu}p^{\mu}$ and $L^{+^{\prime}\mu}=p^{\mu}$ as computed directly from
Table 1.

The simplified version of the BRST gauge symmetry of \cite{2tbrst2006} was
called 2T-gauge-symmetry in \cite{2T SM}, and is given by $\delta_{\Lambda
}\Phi\left(  X\right)  =X^{2}\Lambda\left(  X\right)  .$ In the simplified
version it is useful to think of $\Phi\left(  X\right)  $ in the form
$\Phi\left(  X\right)  =\Phi_{0}\left(  X\right)  +X^{2}\tilde{\Phi}\left(
X\right)  ,$ where $\Phi_{0}\equiv\left[  \Phi\left(  X\right)  \right]
_{X^{2}=0}$ is defined as $\Phi\left(  X\right)  $ evaluated at $X^{2}=0,$ and
$X^{2}\tilde{\Phi}\left(  X\right)  \equiv\Phi\left(  X\right)  -\Phi
_{0}\left(  X\right)  $ is the remainder. According to the gauge
transformation, we see that the remainder $\tilde{\Phi}$ is gauge
freedom\footnote{In this simplified form both $\Lambda\left(  X\right)  $ and
$\tilde{\Phi}\left(  X\right)  $ are a priori taken as homogeneous fields that
satisfy the homogeneity conditions $\left(  X\cdot\partial+\frac{d+2}%
{2}\right)  \tilde{\Phi}\left(  X\right)  =0,$ and similarly for
$\Lambda\left(  X\right)  .$ There is a more complete, but more elaborate form
of the gauge transformation $\delta_{\Lambda}\Phi=\Lambda_{0}+X^{2}\Lambda
_{1}$, with a relation between $\Lambda_{0}$ and $\Lambda_{1},$ that leaves
the same action invariant. With the stronger form of the gauge symmetry the
homogeneity restriction is lifted so that $\tilde{\Phi}\left(  X\right)  $ is
arbitrary. However, using the stronger form we can choose an intermediate
gauge that makes $\tilde{\Phi}\left(  X\right)  $ homogeneous as above, or
gauge fix it to zero directly. \label{homogeneous}}. It is sufficient to
consider the simplified 2T-gauge-symmetry to uniquely determine the form of
the action given above. This gauge symmetry automatically prevents the
appearance of any other power of the field in the potential $V\left(
\Phi\right)  $, including a quadratic mass term\footnote{However, if there are
several fields, their products may appear as long as the total scale dimension
is $d,$ to be cancelled by the scale of the volume element $\left(
d^{d+2}X\right)  \delta\left(  X^{2}\right)  .$}. \ As a consequence, the
$d+2$ theory cannot have a mass term and is invariant under global scale
transformations $\Phi^{\prime}\left(  X\right)  =e^{a\left(  d-2\right)
/2}\Phi\left(  e^{a}X\right)  .$ This scale transformation in $d+2$ dimensions
is separate from the manisfest global symmetry SO$\left(  d,2\right)  .$ Note
that SO$\left(  d,2\right)  $ includes a transformation with generator
$D\equiv L^{+^{\prime}-^{\prime}}$ which turns into a scale transformation in
the lower dimension $x^{\mu}$ when SO$\left(  d,2\right)  $ metamorphoses into
conformal symmetry in the massless particle gauge of Table 1.

Varying the action gives the Euler-Lagrange equation of motion
\begin{equation}
\delta\left(  X^{2}\right)  \left[  \partial^{2}\Phi-V^{\prime}\left(
\Phi\right)  \right]  +2\delta^{\prime}\left(  X^{2}\right)  \left(
X\cdot\partial+\frac{d-2}{2}\right)  \Phi=0.
\end{equation}
This results in two independent equations of motion as coefficients of
$\delta\left(  X^{2}\right)  $ and $\delta^{\prime}\left(  X^{2}\right)  $%
\begin{gather}
\left[  \left(  X\cdot\partial+\frac{d-2}{2}\right)  \Phi_{0}\right]
_{X^{2}=0}=0\\
\left[  \partial^{2}\Phi_{0}-V^{\prime}\left(  \Phi_{0}\right)  -2\left(
X\cdot\partial+\frac{d+2}{2}\right)  \tilde{\Phi}\right]  _{X^{2}=0}=0,\text{
}%
\end{gather}
where we take into account that the equation $\delta\left(  X^{2}\right)
F\left(  X\right)  +\delta^{\prime}\left(  X^{2}\right)  G\left(  X\right)
=0$ has the general solution $\left(  G_{0}\right)  _{X^{2}=0}=0$ and $\left(
F-\tilde{G}\right)  _{X^{2}=0}=0$, where $G\left(  X\right)  =G_{0}\left(
X\right)  +X^{2}\tilde{G}\left(  X\right)  $. Using the gauge symmetry we can
choose the gauge $\tilde{\Phi}=0$, or else impose the homogeneity condition
described in footnote (\ref{homogeneous}). Then the equations can be rewritten
more simply as%
\begin{equation}
\left[  \partial^{2}\Phi-V^{\prime}\left(  \Phi\right)  \right]  _{X^{2}%
=0}=0,\text{ and }\left[  \left(  X\cdot\partial+\frac{d-2}{2}\right)
\Phi\right]  _{X^{2}=0}=0. \label{scalar field eom}%
\end{equation}
These, together with $X^{2}=0,$ are the three 2T equations of motion for a
scalar field that correspond to the three Sp$\left(  2,R\right)  $
constraints, including interactions. \ They were originally found in
\cite{Dirac}\cite{kastrup}\cite{2T field paper} at the level of equations of
motion\footnote{The Sp$\left(  2,R\right)  $ point of view developed as an
independent fundamental principle that coincided with Dirac's approach in this
case. More generally the Sp$\left(  2,R\right)  $ principle is extended with
spin, supersymmetry and all possible background fields
\cite{2tgravembackgrounds}\cite{2tbacgrounds} and can also apply to p-branes.
The Sp$\left(  2,R\right)  $ point of view has also been crucial to recognize
the important consequence of 2T-physics, that there are many gauge choices
which lead to many 1T-physics systems, all unified by Sp$\left(  2,R\right)  $
dualities.}, but now we can derive them from an action principle which is
needed, among other things, for the field quantization of the theory.

\section{Emergent $(d-1)+1$ field theory in flat spacetime gauge
\label{flat field}}

Gauge fixing can now be applied either to the action $\left(
\ref{scalar field action}\right)  $ or to the equations of motion $\left(
\ref{scalar field eom}\right)  $. \ We will now summarize how this is done
\cite{2T field paper}\cite{2T SM} in the \textquotedblleft massless
relativistic particle\textquotedblright\ gauge of Table 1. A class of other
gauges will be discussed in the next section.

In 2T field theory we do not choose a gauge for $X^{M}$ like we do for the
worldline theory $X^{M}\left(  \tau\right)  .$ But instead we parameterize
$X^{M}$ in a form that is parallel to the various gauge choices in Tables 1,2.
Thus, corresponding to the "massless relativistic particle" gauge of Table 1,
we parameterize $X^{M}$ as follows%
\begin{equation}
X^{+\prime}=\kappa,\;X^{-\prime}=\kappa\lambda,\;X^{\mu}=\kappa x^{\mu}.
\label{massless}%
\end{equation}
With this parametrization we see that%
\begin{align}
X^{2}  &  =\kappa^{2}\left(  -2\lambda+x^{2}\right)  ,\\
\left(  d^{\left(  d+2\right)  }X\right)  \delta\left(  X^{2}\right)   &
=\frac{1}{2}\kappa^{d-1}d\kappa d\lambda d^{d}x~\delta\left(  \lambda
-\frac{x^{2}}{2}\right)  . \label{vol}%
\end{align}
Computing the derivatives $\frac{\partial}{\partial X^{M}}$ of the field
$\Phi\left(  X\right)  =\Phi\left(  \kappa,\lambda,x^{\mu}\right)  $ via the
chain rule gives%
\begin{equation}
\frac{\partial\Phi}{\partial X^{\mu}}=\frac{1}{\kappa}\frac{\partial\Phi
}{\partial x^{\mu}},\;\;\frac{\partial\Phi}{\partial X^{-\prime}}\ =\frac
{1}{\kappa}\frac{\partial\Phi}{\partial\lambda},\;\;\frac{\partial\Phi
}{\partial X^{+\prime}}=\frac{1}{\kappa}\left(  \kappa\frac{\partial\Phi
}{\partial\kappa}-\lambda\frac{\partial\Phi}{\partial\lambda}-x^{\mu}%
\frac{\partial\Phi}{\partial x^{\mu}}\right)  .
\end{equation}
This leads to $X^{M}\partial_{M}\Phi\left(  X\right)  =\kappa\frac{\partial
}{\partial\kappa}\Phi\left(  \kappa,\lambda,x^{\mu}\right)  $ and to the
following form of the Laplacian%
\begin{equation}
\partial^{M}\partial_{M}\Phi=\frac{1}{\kappa^{2}}\left[  \left(
\frac{\partial}{\partial x^{\mu}}+x_{\mu}\frac{\partial}{\partial\lambda
}\right)  ^{2}-\left(  2\kappa\frac{\partial}{\partial\kappa}+d-2\right)
\frac{\partial}{\partial\lambda}+(2\lambda-x^{2})\left(  \frac{\partial
}{\partial\lambda}\right)  ^{2}\right]  \Phi. \label{dynamics}%
\end{equation}

We now solve explicitly the equations that follow from $\delta\left(
X^{2}\right)  $ and $\left(  X\cdot\partial+\frac{d-2}{2}\right)  \Phi\left(
X\right)  =0,$ and determine the field configuration that obeys these
\textit{kinematic} conditions, leaving the dynamics of Eq.(\ref{dynamics}) for
later. This is the quantum analog of solving explicitly two out of the three
constraints $X^{2}=X\cdot P=0$ in the worldline theory at the classical level
as displayed in the tables.

We begin by using the 2T-gauge-symmetry which allows us to first choose the
gauge in which the remainder $\tilde{\Phi}\left(  X\right)  $ in the field
\begin{equation}
\Phi\left(  X\right)  =\Phi_{0}\left(  X\right)  +X^{2}\tilde{\Phi}\left(
X\right)  =\Phi_{0}\left(  \kappa,x^{\mu}\right)  -2\kappa^{2}\left(
\lambda-\frac{x^{2}}{2}\right)  \tilde{\Phi}\left(  \kappa,\lambda,x^{\mu
}\right)
\end{equation}
vanishes with a gauge choice $\tilde{\Phi}\left(  X\right)  =0.$ The field
$\Phi_{0}\left(  X\right)  $ is independent of $X^{2}$ by definition, and
therefore it is also independent of $\lambda$ (consider the series expansion
in powers of $\lambda-\frac{q^{2}\left(  x\right)  }{2}$). Hence, in this
gauge we have $\frac{\partial\Phi}{\partial\lambda}=0.$ This is the analog in
the worldline theory to using the gauge $P^{+^{\prime}}\left(  \tau\right)
=0$ of Table 1, whose quantum equivalent is$\;P^{+^{\prime}}\Phi
=-i\frac{\partial\Phi}{\partial X^{-\prime}}\ =-i\frac{1}{\kappa}%
\frac{\partial\Phi}{\partial\lambda}=0.$ With this, all dependence on
$\lambda$ in the field has disappeared and the remaining field takes the
$\lambda$ independent form $\Phi\left(  \kappa,x^{\mu}\right)  $. Hence
$\lambda$ appears only in the volume element (\ref{vol}) and can be integrated out.

Next we solve the kinematic equation $\left(  X\cdot\partial+\frac{d-2}%
{2}\right)  \Phi\left(  X\right)  =\left(  \kappa\frac{\partial}%
{\partial\kappa}+\frac{d-2}{2}\right)  \Phi\left(  \kappa,\lambda,x^{\mu
}\right)  =0$, which results in the solution%
\begin{equation}
\Phi\left(  \kappa,x^{\mu}\right)  =\kappa^{-\frac{d-2}{2}}\phi\left(  x^{\mu
}\right)
\end{equation}
with an arbitrary $\phi\left(  x\right)  .$ Inserting this into the action
(\ref{scalar field action}), and using the now greatly simplified Laplacian%
\begin{equation}
\partial^{M}\partial_{M}\Phi=\kappa^{-\frac{d+2}{2}}\frac{\partial^{2}%
\phi\left(  x\right)  }{\partial x^{\mu}\partial x_{\mu}},
\end{equation}
we obtain a reduced action in $\left(  d-1\right)  +1$ dimensions for the
interacting field $\phi\left(  x^{\mu}\right)  $
\begin{align}
S\left(  \Phi\right)   &  =Z\int\left(  \kappa^{d+1}d\kappa d\lambda
d^{d}x\right)  \frac{1}{2\kappa^{2}}\delta\left(  \lambda-\frac{x^{2}}%
{2}\right) \label{derive}\\
&  \times\left[  \frac{1}{2}\left(  \kappa^{-\frac{d-2}{2}}\phi\right)
\left(  \kappa^{-\frac{d+2}{2}}\frac{\partial^{2}\phi}{\partial x^{\mu
}\partial x_{\mu}}\right)  -\gamma\frac{d-2}{2d}\left(  \kappa^{-\frac{d-2}%
{2}}\phi\right)  ^{\frac{2d}{d-2}}\right]  ,\nonumber\\
&  =\left(  Z\int\frac{d\kappa}{2\kappa}\right)  \int d^{d}x\left[  \frac
{1}{2}\phi\frac{\partial^{2}\phi}{\partial x^{\mu}\partial x_{\mu}}%
-\gamma\frac{d-2}{2d}\phi^{\frac{2d}{d-2}}\right]  ,
\end{align}
The overall factor is normalized by choosing $Z\int\frac{d\kappa}{2\kappa}=1$.
\ We then obtain the reduced action after an integration by parts%
\begin{equation}
S\left(  \phi\right)  =\int d^{d}x\left[  -\frac{1}{2}\frac{\partial\phi
}{\partial x^{\mu}}\frac{\partial\phi}{\partial x_{\mu}}-\gamma\frac{d-2}%
{2d}\phi^{\frac{2d}{d-2}}\right]  . \label{emergentmassless}%
\end{equation}
From here we can proceed to derive the equation of motion for $\phi\left(
x\right)  $ from this action or from the original field equation
(\ref{scalar field eom}) with equivalent results.

The 1T-physics interpretation of the emergent action (\ref{emergentmassless})
is the standard Klein-Gordon massless scalar with a scale invariant
$\phi^{\frac{2d}{d-2}}$ interaction. It is well known (at least in $d=4,$
$\phi^{4}$ theory) that this action has a conformal SO$\left(  d,2\right)  $
symmetry at the classical field theory level. Clearly this SO$\left(
d,2\right)  $ hidden conformal symmetry \cite{Dirac} is nothing but the
manifest Lorentz symmetry of the original action (\ref{scalar field action}),
and is one of the indications that the underlying theory is indeed a theory in
$d+2$ dimensions.

By studying other gauges, such as those listed in Tables 1,2, it is clear that
this message of $d+2$ dimensions with a hidden SO$\left(  d,2\right)  $
symmetry (whose interpretation is different from conformal symmetry) will be
repeated in any other gauge. Furthermore, the emergent field theories in
different gauges will correspond to different 1T-physics interpretations of
the same $d+2$ system (\ref{scalar field eom}), but with predicted field
theoretic duality relations among themselves. This is one of the non-trivial
outputs of 2T-physics, as explored partially in the following sections, for
which 1T-physics gives no clues at all.

\section{Emergent $(d-1)+1$ field theory in curved spacetime gauges
\label{main part}}

We will now discuss a family of gauge choices of the 2T-physics field theory
(\ref{scalar field action}) leading to the following Klein-Gordon field theory
for an interacting scalar field in various curved spacetime backgrounds%
\begin{equation}
S\left(  \phi,g\right)  =\int d^{d}x~\sqrt{-g}\left[  -\frac{1}{2}g^{\mu\nu
}\left(  x\right)  \frac{\partial\phi}{\partial x^{\mu}}\frac{\partial\phi
}{\partial x^{\nu}}-\frac{d-2}{8\left(  d-1\right)  }R\left(  g\right)
\phi^{2}-\gamma\frac{d-2}{2d}\phi^{\frac{2d}{d-2}}\right]  .
\label{confscalar}%
\end{equation}
The explicit form of $R\left(  g\right)  $ for all the metrics listed in
(\ref{list}) are given in Eq.(\ref{table3}). The special coefficient in front
of the curvature term $R\left(  g\right)  $ indicates that this system is
recognized as the conformal scalar in some special backgrounds. Its properties
will be discussed in the following section. The resulting class of special
metrics $g_{\mu\nu}\left(  x\right)  $ consists of all possible conformally
flat metrics that can be written as in Eq.(\ref{confflat}). Among these we
note some interesting cases including those given in the list (\ref{list}). In
particular, when the spacetime is such that $R\left(  g\right)  $ is a
constant (example AdS$_{d-n}\times$S$^{n}$, maximally symmetric space, etc.),
then that term is similar to a mass term. We will derive $S\left(
\phi,g\right)  $ for all $g_{\mu\nu}$ in our list (\ref{list}) by starting
from the 2T field theory (\ref{scalar field action}) and treating the field
theoretic equivalent of the general conformal gauge given in Table 1 for the
worldline theory.

This family of field theoretic gauges does not include the gauge choices in
which some components of $X^{M}\left(  x,p\right)  $ on the worldline depend
both on $x^{\mu}$ and $p^{\mu},$ as seen for some items listed in Tables 1,2,
including the massive relativistic, non-relativistic, H-atom gauges, and more.
A field theoretic study of the non-relativistic massive particle gauge\ can be
found in \cite{2T field paper}. We will return to gauges with nontrivial
momentum dependence in $X^{M}\left(  x,p\right)  $ in future research with the
aim of studying a possible origin for mass in the 2T-physics context.

In this paper we will see that there is hidden SO$\left(  d,2\right)  $
invariance in each curved space theory under a non-linear action of a deformed
conformal group as a manifestation of the original global SO$(d,2)$ symmetry
of the parent theory (\ref{scalar field action}). \ Furthermore, the different
emergent field theories in $\left(  d-1\right)  +1$ dimensions distinguished
by their background metrics will be related to each other by the explicit
duality transformations given in the next section.

To derive these claims, we work in the light cone basis, \ $M=\left(
+^{\prime},-^{\prime},m\right)  $, and parameterize $X^{M}\left(
\kappa,\lambda,x^{\mu}\right)  $ as in (\ref{massless}), in parallel to the
general conformal gauge in Table 1. This has the following form%
\begin{equation}
X^{M}\left(  \kappa,\lambda,x^{\mu}\right)  =\kappa e^{\sigma\left(  x\right)
}\left(  \overset{+^{\prime}}{1},~\overset{-^{\prime}}{\lambda},~\overset
{m}{q^{m}\left(  x\right)  }\right)  \label{parametrization}%
\end{equation}
Solving for $\kappa$, $\lambda$ and $q^{m}\left(  x\right)  $, in terms of
$X^{\pm^{\prime}},X^{m}$ we get the inverse parametrization%
\begin{equation}
\kappa=e^{-\sigma\left(  x\right)  }X^{+\prime},\;\;\lambda=\frac{X^{-\prime}%
}{X^{+\prime}},\;\;q^{m}\left(  x\right)  =\frac{X^{m}}{X^{+\prime}}.
\label{q}%
\end{equation}
From $q^{m}\left(  x\right)  =\frac{X^{m}}{X^{+\prime}}$ we solve for $x^{\mu
}=f^{\mu}\left(  \frac{X^{m}}{X^{+\prime}}\right)  ,$ where $f^{\mu}\left(
q^{m}\right)  $ is the inverse map of $q^{m}\left(  x^{\mu}\right)  .$ This
inverse map is inserted in $\sigma\left(  x\right)  =$ $\sigma\left(  f^{\mu
}\left(  \frac{X^{m}}{X^{+\prime}}\right)  \right)  $ in Eq.(\ref{q}) to
complete the full solution of $\kappa=X^{+\prime}\exp\left(  -\sigma\left(
f^{\mu}\left(  \frac{X^{m}}{X^{+\prime}}\right)  \right)  \right)  $ in terms
of $X^{\pm^{\prime}},X^{m}.$ The field $\Phi\left(  X\right)  =\Phi\left(
\kappa,\lambda,x^{\mu}\right)  $ is now considered a function of
$\kappa,\lambda,x^{\mu}.$ The $\left(  d-1\right)  +1$ spacetime $x^{\mu}$ has
been embedded in $d+2$ dimensions in different forms that vary as the
functions $q^{m}\left(  x\right)  $ and $\sigma\left(  x\right)  $ change (see
examples in Tables 1,2).

With this parametrization we see that%
\begin{align}
X^{2}  &  =\left(  \kappa e^{\sigma}\right)  ^{2}\left(  -2\lambda
+q^{2}\left(  x\right)  \right)  ,\\
\left(  d^{\left(  d+2\right)  }X\right)  \delta\left(  X^{2}\right)   &
=\frac{1}{2}\kappa^{d-1}\det\left(  e_{\mu}^{m}\left(  x\right)  \right)
~d\kappa d\lambda d^{d}x~\delta\left(  \lambda-\frac{q^{2}\left(  x\right)
}{2}\right)  . \label{volume}%
\end{align}
Here we have taken into account the Jacobian
\begin{equation}
J\left(  \frac{X^{+\prime},X^{-\prime},X^{m}}{\kappa,\lambda,x^{\mu}}\right)
=\kappa^{d+1}e^{\left(  d+2\right)  \sigma}\det\left(  \partial_{\mu}%
q^{m}\right)  =\kappa^{d+1}e^{2\sigma}\det\left(  e_{\mu}^{m}\left(  x\right)
\right)  ,
\end{equation}
where the vielbein
\begin{equation}
e_{\mu}^{m}\left(  x\right)  =e^{\sigma\left(  x\right)  }\frac{\partial
q^{m}\left(  x\right)  }{\partial x^{\mu}} \label{vilebein}%
\end{equation}
is the same as the one that emerged in the worldline theory in
Eq.(\ref{confflat}).

Using the chain rule, we can then compute the partial derivatives
$\frac{\partial\Phi}{\partial X^{M}}$ in terms of $\frac{\partial\Phi
}{\partial\kappa},\frac{\partial\Phi}{\partial\lambda},\frac{\partial\Phi
}{\partial x^{\mu}}.$ The result is%
\begin{align}
\frac{\partial\Phi}{\partial X^{-\prime}}  &  =\frac{1}{\kappa}e^{-\sigma
}\frac{\partial\Phi}{\partial\lambda}\label{chain1}\\
\frac{\partial\Phi}{\partial X^{m}}  &  =\frac{1}{\kappa}\left(  -e_{m}^{\mu
}\partial_{\mu}\sigma~\kappa\frac{\partial\Phi}{\partial\kappa}+e_{m}^{\mu
}\partial_{\mu}\Phi\right) \label{chain2}\\
\frac{\partial\Phi}{\partial X^{+\prime}}  &  =\frac{1}{\kappa}\left(  \left[
e^{-\sigma}+q^{m}e_{m}^{\mu}\partial_{\mu}\sigma\right]  ~\kappa\frac
{\partial\Phi}{\partial\kappa}-e^{-\sigma}\lambda\frac{\partial\Phi}%
{\partial\lambda}-q^{m}e_{m}^{\mu}\partial_{\mu}\Phi\right)  \label{chain3}%
\end{align}
Here $e_{m}^{\mu}\left(  x\right)  $ is the inverse vielbein which can also be
written as $e_{m}^{\mu}\left(  x\right)  =e^{-\sigma\left(  x\right)  }%
\frac{\partial x^{\mu}}{\partial q^{m}}=e^{-\sigma\left(  x\right)  }%
\frac{\partial f^{\mu}\left(  q\right)  }{\partial q^{m}}\left(  x\right)  ,$
where $x^{\mu}=f^{\mu}\left(  q\right)  $ is the inverse map discussed
following Eq.(\ref{q}). This is verified by using the chain rule $e_{\nu}%
^{~m}\left(  x\right)  e_{m}^{\mu}\left(  x\right)  =e^{\sigma\left(
x\right)  }\frac{\partial q^{m}}{\partial x^{\nu}}e^{-\sigma\left(  x\right)
}\frac{\partial x^{\mu}}{\partial q^{m}}=\frac{\partial x^{\mu}}{\partial
x^{\nu}}=\delta_{\nu}^{~\mu}.$ We note that the dimension operator takes a
simple form $\kappa\frac{\partial}{\partial\kappa}$
\begin{equation}
X\cdot\partial\Phi=-X^{+^{\prime}}\frac{\partial\Phi}{\partial X^{-\prime}%
}-X^{-^{\prime}}\frac{\partial\Phi}{\partial X^{+\prime}}+X^{m}\frac
{\partial\Phi}{\partial X^{m}}=\kappa\frac{\partial\Phi}{\partial\kappa}.
\end{equation}

We are now ready to choose gauges and solve the two kinematic equations
following in the footsteps of the previous section. We begin by using the
2T-gauge-symmetry which allows us to first choose the gauge in which the
remainder $\tilde{\Phi}\left(  X\right)  $ in the field
\begin{equation}
\Phi\left(  X\right)  =\Phi_{0}\left(  X\right)  +X^{2}\tilde{\Phi}\left(
X\right)  =\Phi_{0}\left(  \kappa,x^{\mu}\right)  -2\left(  \kappa e^{\sigma
}\right)  ^{2}\left(  \lambda-\frac{q^{2}\left(  x\right)  }{2}\right)
\tilde{\Phi}\left(  \kappa,\lambda,x^{\mu}\right)
\end{equation}
vanishes with a gauge choice $\tilde{\Phi}\left(  X\right)  =0.$ The field
$\Phi_{0}\left(  X\right)  $ is independent of $X^{2}$ by definition, and
therefore it is also independent of $\lambda$ (consider the series expansion
in powers of $\lambda-\frac{q^{2}\left(  x\right)  }{2}$). Hence, in this
gauge we have $\frac{\partial\Phi}{\partial\lambda}=0$ so that the remaining
field takes the $\lambda$ independent form $\Phi\left(  \kappa,x^{\mu}\right)
$ everywhere in the action (\ref{scalar field action}). Hence $\lambda$
appears only in the volume element (\ref{volume}) and can be integrated out.
This is the analog in the worldline theory to using the gauge $P^{+^{\prime}%
}\left(  \tau\right)  =0$ of Table 1, whose quantum equivalent
is$\;P^{+^{\prime}}\Phi=-i\frac{\partial\Phi}{\partial X^{-\prime}}%
\ =-i\frac{1}{\kappa}e^{-\sigma}\frac{\partial\Phi}{\partial\lambda}=0.$ Next
we solve the kinematic equation in Eq.(\ref{scalar field eom}) $\left(
X\cdot\partial+\frac{d-2}{2}\right)  \Phi\left(  X\right)  =\left(
\kappa\frac{\partial}{\partial\kappa}+\frac{d-2}{2}\right)  \Phi\left(
\kappa,\lambda,x^{\mu}\right)  =0$, which results in the solution%
\begin{equation}
\Phi=\kappa^{-\frac{d-2}{2}}\phi\left(  x^{\mu}\right)  . \label{phi}%
\end{equation}

So far we have reduced the field to $\left(  d-1\right)  +1$ dimensions by
solving the kinematic constraints in a convenient gauge. We are now ready to
analyze the dynamics satisfied by the field $\phi\left(  x^{\mu}\right)  $. We
compute the Laplacian $\partial^{M}\partial_{M}\Phi=-2\frac{\partial}{\partial
X^{+\prime}}\frac{\partial}{\partial X^{-\prime}}\Phi+\eta^{mn}\frac{\partial
}{\partial X^{m}}\frac{\partial}{\partial X^{n}}\Phi$ by using
Eqs.(\ref{chain1}-\ref{chain3}). Recalling to drop all terms containing
$\frac{\partial\Phi}{\partial\lambda}=0$, we have%
\begin{align}
\partial^{M}\partial_{M}\Phi &  =0+\eta^{mn}\frac{\partial}{\partial X^{m}%
}\frac{\partial}{\partial X^{n}}\Phi\\
&  =\eta^{mn}\left(  -e_{m}^{\mu}\partial_{\mu}\sigma~\frac{\partial}%
{\partial\kappa}+\frac{1}{\kappa}e_{m}^{\mu}\partial_{\mu}\right)  \left(
-e_{n}^{\nu}\partial_{\nu}\sigma~\frac{\partial\Phi}{\partial\kappa}+\frac
{1}{\kappa}e_{n}^{\nu}\partial_{\nu}\Phi\right) \nonumber
\end{align}
After inserting $\Phi\left(  \kappa,x^{\mu}\right)  =\kappa^{-\frac{d-2}{2}%
}\phi\left(  x^{\mu}\right)  $ and some calculation, this takes the form%
\begin{equation}
\partial^{M}\partial_{M}\Phi=\kappa^{-\frac{d+2}{2}}\left\{
\begin{array}
[c]{c}%
g^{\mu\nu}\partial_{\mu}\partial_{\nu}\phi+\left(  e^{n\mu}\partial_{\mu}%
e_{n}^{\nu}+\left(  d-1\right)  g^{\mu\nu}\partial_{\mu}\sigma\right)
\partial_{\nu}\phi\\
+\left[  \frac{(d-2)d}{4}g^{\mu\nu}\partial_{\mu}\sigma\partial_{\nu}%
\sigma+\frac{d-2}{2}\left(  g^{\mu\nu}\partial_{\mu}\partial_{\nu}%
\sigma+e^{n\mu}\partial_{\mu}e_{n}^{\nu}\partial_{\nu}\sigma\right)  \right]
\phi
\end{array}
\right\}
\end{equation}
where we have defined the metric
\begin{equation}
g^{\mu\nu}\left(  x\right)  \equiv\eta^{mn}e_{m}^{\mu}\left(  x\right)
e_{n}^{\nu}\left(  x\right)  =e^{2\sigma\left(  x\right)  }\eta_{mn}%
\partial_{\mu}q^{m}\left(  x\right)  \partial_{\nu}q^{n}\left(  x\right)  ,
\label{metric}%
\end{equation}
This metric is conformally flat. The expression can be further simplified to
take the form of the Laplacian operator for the metric $g^{\mu\nu}\left(
x\right)  $ with an additional curvature term\footnote{To bring the Laplacian
to this form we have used the following identities $\sqrt{-g}=e=\det\left(
e_{\mu}^{m}\right)  ,$ and $\partial_{\mu}g^{\mu\nu}=\partial_{\mu}\left(
e^{n\mu}e_{n}^{\nu}\right)  =e^{n\mu}\partial_{\mu}e_{n}^{\nu}+e_{n}^{\nu
}\partial_{\mu}e^{n\mu},$ and%
\begin{equation}
\frac{1}{e}\partial^{\nu}e=\partial^{\nu}\ln\left[  \det\left(  e_{\mu}%
^{k}\right)  \right]  =\partial^{\nu}\ln\left[  e^{tr\ln\left(  e_{\mu}%
^{k}\right)  }\right]  =tr\left[  \partial^{\nu}\ln\left(  e_{\mu}^{k}\right)
\right]  =tr\left[  e_{k}^{\lambda}\partial^{\nu}e_{\mu}^{k}\right]
=e_{k}^{\mu}\partial^{\nu}e_{\mu}^{k}%
\end{equation}
which gives $\frac{1}{e}\partial^{\nu}e+e_{n}^{\nu}\partial_{\mu}e^{n\mu
}=e_{k}^{\mu}\partial^{\nu}e_{\mu}^{k}+e_{n}^{\nu}\partial_{\mu}e^{n\mu
}=\left(  d-1\right)  \partial^{\nu}\sigma.$}%
\begin{equation}
\partial^{M}\partial_{M}\Phi=\kappa^{-\frac{d+2}{2}}\left\{  \frac{1}%
{\sqrt{-g}}\partial_{\mu}\left(  \sqrt{-g}g^{\mu\nu}\partial_{\nu}\phi\right)
-\frac{d-2}{4\left(  d-1\right)  }R\left(  g\right)  \phi\right\}  .
\label{laplacian}%
\end{equation}
The curvature scalar $R\left(  g\right)  $ for the given metric in
Eq.(\ref{metric}) is computed in Appendix \ref{Riemann Ricci}%
\begin{equation}
R\left(  g\right)  =\left(  1-d\right)  \left[  dg^{\mu\nu}\partial_{\mu
}\sigma\partial_{\nu}\sigma+2g^{\mu\nu}\partial_{\mu}\partial_{\nu}%
\sigma+2e^{n\mu}\partial_{\mu}e_{n}^{\nu}\partial_{\nu}\sigma\right]
\end{equation}

Inserting the reduced forms for the field (\ref{phi}), the Laplacian
(\ref{laplacian}), and the volume element (\ref{volume}) into the $d+2$
dimensional action (\ref{scalar field action}), and repeating the steps of the
reduction procedure similar to Eq.(\ref{derive}), we finally derive the
emergent action for the interacting conformal scalar given in
Eq.(\ref{confscalar}).

In the following two sections it will be shown that two such emergent actions,
with different background metrics $g_{\mu\nu}\left(  x\right)  $ and
$\tilde{g}_{\mu\nu}\left(  x\right)  $ such as those listed in (\ref{list}),
are dual to each other, while for each fixed metric $g_{\mu\nu}\left(
x\right)  $ of this type the action is invariant under a hidden SO$\left(
d,2\right)  $ global symmetry. These properties emanate, of course, from the
the original action, and are indications of the underlying spacetime in $d+2$ dimensions.

\section{Dualities\label{conf KG eq}}

Given any two metrics $g_{\mu\nu}\left(  x\right)  $ and $\tilde{g}_{\mu\nu
}\left(  x\right)  $ in $\left(  d-1\right)  +1$ dimensions, built from
$\left(  \sigma\left(  x^{\mu}\right)  ,q^{m}\left(  x^{\mu}\right)  \right)
$ or $\left(  \tilde{\sigma}\left(  x^{\mu}\right)  ,\tilde{q}^{m}\left(
x^{\mu}\right)  \right)  $ respectively as in Eq.(\ref{metric}), we have two
KG field theories that are considered theories in different fixed background
spacetimes from the point of view of 1T-physics. However, since we obtained
them by gauge fixing the same parent 2T field theory $S\left(  \Phi\right)  $
in $d+2$ dimensions we have the 2T-physics prediction that they are in some
sense the same action%
\begin{equation}
S\left(  \phi,g_{\mu\nu}\right)  =S\left(  \Phi\right)  =S\left(  \tilde{\phi
},\tilde{g}_{\mu\nu}\right)  . \label{duality2T}%
\end{equation}
So we expect a duality transformation that relates two different 1T-physics
actions $S\left(  \phi,g_{\mu\nu}\right)  $, $S\left(  \tilde{\phi},\tilde
{g}_{\mu\nu}\right)  ,$ for the classes of \textit{metrics related by
2T-physics }as specified in the previous section. This duality transformation
is constructed explicitly in this section.

In the worldline theory at the classical level the relationship between any
two gauges in Tables 1,2 is given by an Sp$\left(  2,R\right)  $ gauge
transformation in \textit{phase space}%
\begin{equation}
\left(
\begin{array}
[c]{c}%
\tilde{X}^{M}\left(  \tilde{t},\tilde{H},\mathbf{\tilde{r}},\mathbf{\tilde{p}%
}\right) \\
\tilde{P}_{M}\left(  \tilde{t},\tilde{H},\mathbf{\tilde{r}},\mathbf{\tilde{p}%
}\right)
\end{array}
\right)  =\left(
\begin{array}
[c]{cc}%
a & b\\
c & d
\end{array}
\right)  \left(
\begin{array}
[c]{c}%
X^{M}\left(  t,H,\mathbf{r,p}\right) \\
P_{M}\left(  t,H,\mathbf{r,p}\right)
\end{array}
\right)  \label{sp2r}%
\end{equation}
where the Sp$\left(  2,R\right)  $ parameters $\left(  a,b,c,d\right)  \left(
\tau\right)  $ that are local on the worldline can be written as functions of
either set of canonical variables $\left(  \tilde{t},\tilde{H},\mathbf{\tilde
{r}},\mathbf{\tilde{p}}\right)  \left(  \tau\right)  $ or $\left(
t,H,\mathbf{r,p}\right)  \left(  \tau\right)  .$ What form does this
transformation take in the local field theory in \textit{position space}
alone, and in terms of the dynamical field $\phi\left(  x\right)  $?

For the class of spacetimes we have specialized above the $X^{M}\left(
x^{\mu}\right)  $ is a function of position space only (does not contain
$p^{\mu}$), and therefore the Sp$\left(  2,R\right)  $ gauge transformation
(\ref{sp2r}) is easily rephrased as a transformation of the type%
\begin{equation}
\left(  \tilde{\sigma}\left(  x^{\mu}\right)  ,\tilde{q}^{m}\left(  x^{\mu
}\right)  \right)  \rightarrow\left(  \sigma\left(  x^{\mu}\right)
,q^{m}\left(  x^{\mu}\right)  \right)  \label{special}%
\end{equation}
among the metrics (\ref{metric}) such as those listed in (\ref{list}). This
amounts to general coordinate transformations $x^{\mu}\rightarrow y^{\mu
}\left(  x\right)  $ and local Weyl rescaling of the metric, which we will
implement below in Eqs.(\ref{gal coord transf1}-\ref{Weyl2}) in the field
theory language. It is evident that our approach allows us to contemplate the
more general duality transformation that goes beyond (\ref{special}) and
thereby include in our discussion field theoretic duality transformations to
the other gauge types $X^{M}\left(  t,H,\mathbf{r,p}\right)  $ listed in
Tables 1,2, including the massive relativistic, massive non-relativistic, and
H-atom gauges. However, for simplicity we concentrate in this paper on the
easier case of type (\ref{special}).

The action (\ref{confscalar}) is formally invariant under general coordinate
transformations, but of course, since the metric is not dynamical, the action
is not actually invariant. Instead, the general coordinate transformation
$x^{\mu}\rightarrow y^{\mu}(x)$ of the dynamical field $\phi\left(  x\right)
$ maps the theory into another theory with a new background metric that is
related to the old one by the following transformations%
\begin{align}
\phi\left(  x\right)   &  \rightarrow\tilde{\phi}\left(  x\right)
=\phi\left(  y\left(  x\right)  \right)  ,\label{gal coord transf1}\\
g_{\mu\nu}\left(  x\right)   &  \rightarrow\tilde{g}_{\mu\nu}\left(  x\right)
=\partial_{\mu}y^{\lambda}\left(  x\right)  \partial_{\nu}y^{\sigma}\left(
x\right)  g_{\lambda\sigma}\left(  y\left(  x\right)  \right)  .
\label{gal coord transf2}%
\end{align}
Less familiar is that, thanks to special coefficient of the curvature term,
the action (\ref{confscalar}) is also formally invariant under the following
Weyl rescaling \cite{Weyl weight} of the field $\phi$ and metric $g_{\mu\nu}$
\begin{align}
\phi\left(  x\right)   &  \rightarrow\tilde{\phi}\left(  x\right)
=e^{-\frac{d-2}{2}\lambda\left(  x\right)  }\phi\left(  x\right)
,\label{Weyl}\\
g_{\mu\nu}\left(  x\right)   &  \rightarrow\tilde{g}_{\mu\nu}\left(  x\right)
=e^{2\lambda\left(  x\right)  }g_{\mu\nu}\left(  x\right)  , \label{Weyl2}%
\end{align}
as proven below. The rescaling and general coordinate transformations of the
field $\phi$ are induced through the expression $\Phi\left(  X\right)
=\kappa^{-\frac{d-2}{2}}\phi\left(  x\right)  $ in Eq.(\ref{phi}), and can be
understood as originating from the Sp$\left(  2,R\right)  $ gauge
transformation (\ref{sp2r},\ref{special}) of the variables $\left(
\kappa,\lambda,x^{\mu}\right)  $ defined through (\ref{q}).

Again it should be emphasized that this Weyl rescaling is not an invariance of
the action (\ref{confscalar}) since the metric $g_{\mu\nu}\left(  x\right)  $
is not dynamical. Rather, this is a duality transformation to another theory
with a new background metric $\tilde{g}_{\mu\nu}\left(  x\right)  $. \

The duality under the general coordinate transformation is evident. We will
now prove the duality under the Weyl rescaling (\ref{Weyl},\ref{Weyl2}). For
this, it will be useful to provide the following transformation rules for the
curvature tensors which are well known%

\begin{equation}
\tilde{\Gamma}_{\mu\nu}^{\lambda}=\Gamma_{\mu\nu}^{\lambda}+\delta_{\mu
}^{\lambda}\partial_{\nu}\lambda+\delta_{\nu}^{\lambda}\partial_{\mu}%
\lambda-g_{\mu\nu}g^{\lambda\kappa}\partial_{\kappa}\lambda,
\end{equation}%
\begin{equation}
\tilde{R}_{\text{ \ \ }\kappa\mu\nu}^{\lambda}=R_{\text{ \ \ }\kappa\mu\nu
}^{\lambda}-2\left\{
\begin{array}
[c]{c}%
\left(  \delta_{\lbrack\mu}^{\lambda}\delta_{\nu]}^{\alpha}\delta_{\kappa
}^{\beta}-\tilde{g}_{\kappa\lbrack\mu}\delta_{\nu]}^{\alpha}\tilde{g}%
^{\lambda\beta}\right)  \nabla_{\alpha}\left(  \partial_{\beta}\lambda\right)
\\
-\left(  \delta_{\lbrack\mu}^{\lambda}\delta_{\nu]}^{\alpha}\delta_{\kappa
}^{\beta}-\tilde{g}_{\kappa\lbrack\mu}\delta_{\nu]}^{\alpha}\tilde{g}%
^{\lambda\beta}+\tilde{g}_{\kappa\lbrack\mu}\delta_{\nu]}^{\lambda}\tilde
{g}^{\alpha\beta}\right)  \partial_{\alpha}\lambda\partial_{\beta}\lambda
\end{array}
\right\}  ,
\end{equation}%
\begin{equation}
\tilde{R}_{\mu\nu}=R_{\mu\nu}-\left[  \left(  d-2\right)  \delta_{\mu}%
^{\kappa}\delta_{\nu}^{\lambda}+g_{\mu\nu}g^{\kappa\lambda}\right]
\nabla_{\kappa}\left(  \partial_{\lambda}\lambda\right)  +\left(  d-2\right)
\left(  \delta_{\mu}^{\kappa}\delta_{\nu}^{\lambda}-g_{\mu\nu}g^{\kappa
\lambda}\right)  \partial_{\kappa}\lambda\partial_{\lambda}\lambda,
\end{equation}%
\begin{equation}
\tilde{R}=e^{-2\lambda}\left\{  R-\left(  d-1\right)  \left[  2g^{\mu\nu
}\nabla_{\mu}\left(  \partial_{\nu}\lambda\right)  +\left(  d-2\right)
g^{\mu\nu}\partial_{\mu}\lambda\partial_{\nu}\lambda\right]  \right\}  ,
\end{equation}
where $\nabla_{\mu}$ is the general covariant derivative. It is now easy to
check that the Lagrangian $\mathcal{L}$ is invariant up to a total divergence.
First we substitute for $\tilde{\phi},\tilde{g}_{\mu\nu},\tilde{R}$%

\begin{equation}
\mathcal{\tilde{L}=}e^{\lambda d}\sqrt{-g}\left(
\begin{array}
[c]{c}%
-\frac{1}{2}\left(  e^{-2\lambda}g^{\mu\nu}\right)  \partial_{\mu}\left(
e^{-\frac{d-2}{2}\lambda}\phi\right)  \partial_{\nu}\left(  e^{-\frac{d-2}%
{2}\lambda}\phi\right)  -\gamma\frac{d-2}{2d}e^{-d\lambda}\phi^{\frac{2d}%
{d-2}}\\
-\frac{d-2}{8\left(  d-1\right)  }e^{-2\lambda}\left\{  R-\left(  d-1\right)
\left[  2g^{\mu\nu}\nabla_{\mu}\left(  \partial_{\nu}\lambda\right)  +\left(
d-2\right)  g^{\mu\nu}\partial_{\mu}\lambda\partial_{\nu}\lambda\right]
\right\}  e^{-\left(  d-2\right)  \lambda}\phi^{2}%
\end{array}
\right)
\end{equation}
For infinitesimal $\lambda\left(  x\right)  $ this expression becomes
$\mathcal{\tilde{L}}=\mathcal{L}+\delta_{\lambda}\mathcal{L},$ where%
\begin{equation}
\delta_{\lambda}\mathcal{L=}\partial_{\mu}\left(  \frac{\left(  d-2\right)
}{4}\sqrt{-g}g^{\mu\nu}\partial_{\nu}\lambda\phi^{2}\right)  .
\end{equation}
Since $\delta_{\lambda}\mathcal{L}$ is a total divergence, this proves that
the two actions $S\left(  g,\phi\right)  $ and $S\left(  \tilde{g},\tilde
{\phi}\right)  $ are indeed related to each other by the duality
transformation (\ref{Weyl},\ref{Weyl2}).

Technically, what we have is a family of actions, some of which are listed in
(\ref{list}), that are related by coordinate and Weyl transformations of the
background metrics, and field redefinitions of the dynamical field
$\phi\left(  x\right)  $. \ This corresponds precisely to the duality
predicted by 2T physics in Eq.(\ref{duality2T}). \ Alternatively, these
actions in $\left(  d-1\right)  +1$ dimensions can be taken to be different
parameterizations of the same 2T-physics system (\ref{scalar field action}) in
$d+2$ dimensions, however, these different descriptions do not have the same
1T-physics spacetimes, and therefore they have different 1T-physics interpretations.

\section{SO$\left(  d,2\right)  $ global symmetry and its
generators\label{LMN gen}}

In this section, we will show that the KG field theory of Eq.(\ref{confscalar}%
) has a hidden global SO$\left(  d,2\right)  $ in any of the emergent
background spacetimes such as those given in Eq.(\ref{table3}), and we derive
explicitly the form of the $L^{MN}$ generators.

The original master action (\ref{scalar field action}) that led to the
emergent field theories (\ref{confscalar})\ is manifestly invariant under
SO$\left(  d,2\right)  $ global transformations given by%
\begin{equation}
\delta_{\omega}\Phi\left(  X\right)  =\frac{i}{2}\omega_{MN}L^{MN}\Phi\left(
X\right)  ,\;\;L^{MN}\Phi\left(  X\right)  =-i\left(  X^{M}\frac{\partial
\Phi\left(  X\right)  }{\partial X^{N}}-X^{N}\frac{\partial\Phi\left(
X\right)  }{\partial X^{M}}\right)
\end{equation}
The form of the $L^{MN}$ in the emergent field theories can be obtained as
differential operators from this expression by substituting the
parametrization $X^{M}\left(  \kappa,\lambda,x^{\mu}\right)  =\kappa
e^{\sigma\left(  x\right)  }(\overset{+^{\prime}}{1},~\overset{-^{\prime}%
}{\lambda},~\overset{m}{q^{m}\left(  x\right)  })$ and using the chain rules
Eqs.( \ref{chain1}, \ref{chain2}, \ref{chain3}). Applying these on the gauge
fixed form of the field $\Phi\left(  X\right)  =\kappa^{-\frac{d-2}{2}}%
\phi\left(  x^{\mu}\right)  $ given in (\ref{phi}), this procedure provides an
expression of the form%
\begin{equation}
\delta_{\omega}\Phi\left(  X\right)  =\kappa^{-\frac{d-2}{2}}\frac{i}{2}%
\omega_{MN}\left(  L^{MN}\phi\left(  x^{\mu}\right)  \right)  \equiv
\kappa^{-\frac{d-2}{2}}\delta_{\omega}\phi\left(  x^{\mu}\right)
\end{equation}
where $\delta_{\omega}\phi\left(  x^{\mu}\right)  =\frac{i}{2}\omega
_{MN}L^{MN}\phi\left(  x^{\mu}\right)  $ is now given in terms of a non-linear
differential operator representation of the SO$\left(  d,2\right)  $
generators written in terms of the emergent spacetime in $\left(  d-1\right)
+1$ dimensions. Since the original action is invariant under SO$\left(
d,2\right)  ,$ the emergent action must have a hidden SO$\left(  d,2\right)  $
symmetry under this transformation.

We now give the explicit expression for $L^{MN}\phi\left(  x^{\mu}\right)  .$
Rather than presenting the result of the straightforward computation we have
just outlined above, we provide additional insight by also giving the
following arguments based on dualities which lead to the same explicit form
for $L^{MN}\phi\left(  x^{\mu}\right)  .$

Another way to find the generators $L^{MN}\phi\left(  x^{\mu}\right)  $ is to
relate our conformal KG field $\phi\left(  x\right)  $ to the flat KG field
$\phi_{0}\left(  x\right)  $ by the duality transformation in which we take
$\lambda\left(  x\right)  =\sigma\left(  x\right)  $ and $y^{m}\left(
x\right)  =q^{m}\left(  x\right)  $%
\begin{equation}
\phi\left(  x\right)  =e^{-\frac{d-2}{2}\sigma}\phi_{0}\left(  q\left(
x\right)  \right)  .
\end{equation}
Given that the metric in flat space is $\eta_{mn},$ the combined duality
transformations (\ref{gal coord transf2},\ref{Weyl2}) generate the metric
$g_{\mu\nu}=e^{2\sigma}\partial_{\mu}q^{m}\partial_{\nu}q^{n}\eta_{mn}$.

We can then relate the $L^{MN}$'s in curved space to the ones in flat space
$L_{0}^{MN}$
\begin{equation}
\text{flat:\ }\delta_{\omega}\phi_{0}\left(  q\right)  =\frac{i}{2}\omega
_{MN}L_{0}^{MN}\phi_{0}%
\end{equation}
where the $L_{0}^{MN}$ in flat space were obtained directly from 2T-physics in
\cite{2T basics} as the $L^{MN}=X^{M}P^{N}-X^{N}P^{M}$ specialized to flat
spacetime at the quantum level, and given as
\begin{align}
L_{0}^{+\prime m}  &  =p^{m},\;\;L_{0}^{+\prime-\prime}=\frac{1}{2}\left(
p_{m}q^{m}+q^{m}p_{m}\right)  +i\\
L_{0}^{-\prime m}  &  =\frac{1}{2}q_{k}p^{m}q^{k}-\frac{1}{2}q^{m}p_{k}%
q^{k}-\frac{1}{2}q^{k}p_{k}q^{m}-iq^{m}\\
L_{0}^{mn}  &  =q^{m}p^{n}-q^{n}p^{m}%
\end{align}
These correspond to the well known conformal transformations of a scalar field
in flat spacetime. They are Hermitian relative to the norm defined for the
field $\phi_{0}\left(  x\right)  $ in relativistic field theory \cite{2T
basics}. They close under commutation by using $[q_{m},p^{n}]=i$ and
automatically give a constant value for the quadratic Casimir eigenvalue
$C_{2}=\frac{1}{2}L^{MN}L_{MN}=1-d^{4}/4$ independent of $q,p$, which
corresponds to the singleton.

In the present case, as applied on $\phi_{0}\left(  q\left(  x\right)
\right)  ,$ we must replace the symbols $q^{m},p_{m}=-i\frac{\partial
}{\partial q^{m}}$ (and $p^{m}\equiv\eta^{mn}p_{n}$ and $q_{k}\equiv\eta
_{km}q^{m}$) above with the expressions $q^{m}\left(  x\right)  $ and
$p_{m}=-i\frac{\partial x^{\mu}}{\partial q^{m}}\partial_{\mu}$ which can be
written as (see comments following (\ref{chain3}))
\begin{equation}
p_{m}=-ie^{\sigma\left(  x\right)  }e_{m}^{~\mu}\left(  x\right)
\frac{\partial}{\partial x^{\mu}}.
\end{equation}
These generate a differential operator representation for $L_{0}^{MN}$ in
$x^{\mu}$ rather than $q^{m}$ space.

For the general case we can now write%
\begin{align}
\delta_{\omega}\phi\left(  x\right)   &  =e^{-\frac{d-2}{2}\sigma\left(
x\right)  }\delta_{\omega}\phi_{0}\left(  q\left(  x\right)  \right)
=\frac{i}{2}\omega_{MN}e^{-\frac{d-2}{2}\sigma\left(  x\right)  }L_{0}%
^{MN}\phi_{0}\left(  q\left(  x\right)  \right) \\
&  =\frac{i}{2}\omega_{MN}e^{-\frac{d-2}{2}\sigma\left(  x\right)  }L_{0}%
^{MN}\left(  e^{\frac{d-2}{2}\sigma\left(  x\right)  }\phi\left(  x\right)
\right)  \equiv\frac{i}{2}\omega_{MN}L^{MN}\phi\left(  x\right)
\end{align}
Hence the general $L^{MN}$ is given by the differential operators%
\begin{equation}
L^{MN}=e^{-\frac{d-2}{2}\sigma}L_{0}^{MN}e^{\frac{d-2}{2}\sigma} \label{LMN}%
\end{equation}
The closure of the Lie algebra is evident from the form of $L^{MN}$ as a
similarity transformation and the known closure of $L_{0}^{MN}$ as SO$\left(
d,2\right)  $ generators.

In this form, the generator $L^{MN}$ is presented as a combination of duality
plus conformal transformations in ordinary flat spacetime. That is, we start
with the field $\phi\left(  x\right)  ,$ transform it via duality (Weyl and
general coordinate transformation) to the field in flat space $\phi_{0}\left(
q\right)  ,$ apply ordinary SO$\left(  d,2\right)  $ conformal transformations
in the flat space $q^{m},$ and then apply a duality transformation back to the
field $\phi\left(  x\right)  .$

The parameters in this transformation are only the global parameters of
SO$\left(  d,2\right)  $ in flat spacetime, while here $\sigma\left(
x\right)  $ and $q^{m}\left(  x\right)  $ are not parameters since they define
the metric $g_{\mu\nu}\left(  x\right)  $. We emphasize that this deformed
conformal transformation generates a global SO$\left(  d,2\right)  $ symmetry
of the action $S\left(  g,\phi\right)  $ for each \textit{fixed metric}
$g_{\mu\nu}\left(  x\right)  .$

Performing the SO$\left(  d,2\right)  $ transformation $\delta_{\omega}%
\phi\left(  x\right)  =\frac{i}{2}\omega_{MN}L^{MN}\phi\left(  x\right)  $ one
can now see that $\delta_{\omega}S\left(  g,\phi\right)  =0,$ since the
invariance is true for the flat theory with the Minkowski metric $\eta_{\mu
\nu}$ and we have also shown that the duality transformation relates the flat
and curved theories. Without reference to the flat theory, but only using the
generator $L^{MN}$ above, we see that this is a true invariance of the action
$S\left(  g,\phi\right)  $ since the fixed background metric $g_{\mu\nu
}\left(  x\right)  $ is left unchanged by the SO$\left(  d,2\right)  $
transformation. This hidden global symmetry is nothing but the original global
SO$\left(  d,2\right)  $ symmetry of the action $S\left(  \Phi\right)  $ in
$d+2$ dimensions, and hence it is one of the indications within 1T-physics of
the higher dimensional nature of the underlying system.

\section{Conclusion}

In this paper, we have shown that the conformal scalar propagating in any
conformally flat metric in $\left(  d-1\right)  +1$ dimensions can be obtained
using 2T field theory in flat $d+2$ dimensions. \ The SO$\left(  d,2\right)  $
global symmetry of the 2T theory is realized as the hidden non-linear
invariance of the resulting KG equation under the action of a deformed
conformal group. \ The duality between the different conformal KG equations in
different backgrounds is a first step to demonstrate the use of duality in
1T-physics as emergent from 2T field theory. \ As mentioned in the
introduction, the availability of such dualities is expected to be an
important tool in the study of more complicated cases.

The obvious next step in our investigation is to generalize this paper's
results to the spin-$\frac{1}{2}$ and spin-1 cases, and apply these duality
ideas to the Standard Model as a theory that emerges from 4+2 dimensions
\cite{2T SM}. \ This is rather straightforward and is done in a companion
paper \cite{higher spins}. \

The general setup of 2T physics presented in this paper teaches us that the
particular class of dual theories studied in this paper (which we related to
well known properties of the conformal scalar) is only the most evident sector
of a much larger duality, which would be much harder to notice and, arguably,
impossible to investigate systematically from a strict 1T perspective. \ This
paper is indeed the first step of such a program.

The larger set of dualities already uncovered in the worldline formalism leads
us to expect a similar variety in field theory. \ Of particular interest will
be the extension of the theory to allow gauge choices equivalent to those in
the worldline formalism which involve mixing of $x$ and $p$ (cf. section
\ref{gauge choices}). \ This may result in dualities between local and
non-local field theories at least in some instances. \ It is to be noted that
the appearance of mass in the worldline formalism was related to such gauge
choices. \ This suggests the possibility that mass in field theory might come
as a modulus in the embedding of 3+1 dimensional phase space into 4+2
dimensional phase space.

As in other instances of dualities, in principle the class of dualities
described in this paper, and the more general dualities provided by 2T-physics
can be used as new tools to investigate the properties of the Standard Model,
including QCD. For instance, one could use one form of the 1T-physics action
to learn some non-perturbative information about the other 1T-physics action.

Through the dualities, but especially through the parent 2T theory, we obtain
a new unification of 1T field theories through higher dimensions. This is
quite different from the Kaluza-Klein type ideas since there are no
Kaluza-Klein remnants either in the form of extra fields or in the form of
extra quantum numbers. Instead what we have is a family of dual theories with
a different set of parameters described as the moduli of the metrics (and more
generally masses, couplings, etc.), as seen in Tables 1,2. \

Further research on these topics is warranted and is currently being pursued.

\begin{acknowledgments}
We would like to thank Y.-C. Kuo and B. Orcal for helpful discussions.
\end{acknowledgments}

\appendix{}

\section{Curvature tensors\label{Riemann Ricci}}

In this appendix,we provide the Riemann tensor, Ricci tensor, and Ricci scalar
of the $\left(  d-1\right)  +1$ spacetime. \ As a reminder, we use:
\begin{equation}
e_{\mu}^{m}\equiv e^{\sigma}\partial_{\mu}q^{m},g_{\mu\nu}\equiv\eta
_{mn}e_{\mu}^{m}e_{\nu}^{n}=e^{2\sigma}\partial_{\mu}q^{m}\partial_{\nu}%
q^{n}\eta_{mn}.
\end{equation}
The connection is defined as:%
\begin{equation}
\Gamma_{\mu\nu}^{\rho}=\frac{1}{2}g^{\rho\tau}\left(  \partial_{\mu}g_{\nu
\tau}+\partial_{\nu}g_{\mu\tau}-\partial_{\tau}g_{\mu\nu}\right)
\end{equation}
This gives:%
\begin{equation}
\Gamma_{\mu\nu}^{\rho}=\delta_{\mu}^{\rho}\partial_{\nu}\sigma-g_{\mu\nu
}g^{\rho\lambda}\partial_{\lambda}\sigma+e_{m}^{\rho}\partial_{\mu}e_{\nu}^{m}%
\end{equation}
The Riemann tensor is given by%
\begin{align}
R_{\text{ \ \ }\tau\mu\nu}^{\rho}  &  =\partial_{\mu}\Gamma_{\nu\tau}^{\rho
}-\partial_{\nu}\Gamma_{\mu\tau}^{\rho}+\Gamma_{\mu\lambda}^{\rho}\Gamma
_{\nu\tau}^{\lambda}-\Gamma_{\nu\lambda}^{\rho}\Gamma_{\mu\tau}^{\lambda}\\
&  =\left[
\begin{array}
[c]{c}%
\delta_{\mu}^{\rho}\partial_{\nu}\sigma\partial_{\tau}\sigma-g_{\nu\tau}%
e_{i}^{\rho}\partial_{\mu}e^{i\lambda}\partial_{\lambda}\sigma-g_{\nu\tau
}g^{\rho\lambda}\partial_{\lambda}\partial_{\mu}\sigma\\
-\delta_{\mu}^{\rho}g_{\nu\tau}g^{\alpha\beta}\partial_{\alpha}\sigma
\partial_{\beta}\sigma+\delta_{\nu}^{\rho}\partial_{\tau}\sigma\partial_{\mu
}\sigma\\
+\delta_{\nu}^{\rho}e_{i\tau}\partial_{\mu}e^{i\lambda}\partial_{\lambda
}\sigma+\delta_{\nu}^{\rho}\partial_{\tau}\partial_{\mu}\sigma
\end{array}
\right]  -\left(  \mu\leftrightarrow\nu\right)
\end{align}
While the calculation above is straightforward, it is rather tedious and not
particularly enlightening. The Ricci tensor is easily obtained: $R_{\tau\nu
}=R_{\text{ \ \ }\tau\mu\nu}^{\mu}$%
\begin{equation}%
\begin{array}
[c]{c}%
R_{\tau\nu}=\left(  1-d\right)  g_{\nu\tau}g^{\alpha\beta}\partial_{\alpha
}\sigma\partial_{\beta}\sigma\\
-g_{\nu\tau}e_{i}^{\mu}\partial_{\mu}e^{i\lambda}\partial_{\lambda}%
\sigma-g_{\nu\tau}g^{\alpha\beta}\partial_{\alpha}\partial_{\beta}\sigma\\
(2-d)e_{i\tau}\partial_{\nu}e^{i\lambda}\partial_{\lambda}\sigma
+(2-d)\partial_{\tau}\partial_{\nu}\sigma
\end{array}
\end{equation}

Finally the Ricci scalar is: $R=g^{\tau\nu}R_{\tau\nu}$%
\begin{equation}
R=\left(  1-d\right)  \left[  dg^{\mu\nu}\partial_{\mu}\sigma\partial_{\nu
}\sigma+2g^{\mu\nu}\left(  \partial_{\mu}\partial_{\nu}\sigma\right)  +2e^{\nu
m}\partial_{\nu}e_{m}^{\mu}\partial_{\mu}\sigma\right]  \label{R}%
\end{equation}

\end{document}